\documentclass[pre,twocolumn]{revtex4}
\usepackage{epsfig,graphics,subfigure,amssymb,amsmath,subeqnarray,color,xcolor,amssymb}
\usepackage[colorlinks=true, citecolor=red, linkcolor=blue,urlcolor=blue ]{hyperref}
\usepackage{amsthm}
\usepackage{graphicx}
\usepackage{xcolor}
\usepackage{float}

\usepackage[T1]{fontenc}
\usepackage[utf8]{inputenc}

\begin{document}
\def\u{{\bf u}}\def\n{{\bf n}}\def\f{{\bf f}}
\def\t{{\bf t}}\def\d{{\rm d}}\def\Pe{{\rm Pe}}
\def\r{{\bf r}}\def\x{{\bf x}}
\def\e{{\bf e}}\def\1{{\bf 1}}\def\0{{\bf 0}}
\def\p{{\partial}}
\def\C{{\boldsymbol C}}\def\D{{\boldsymbol D}}
\def\x{{\boldsymbol x}}
\def\v{\vspace {1cm}}
\def\textheight{25cm}

\newcommand{\bb}[1]{{[#1]}}
\newcommand{\rr}[1]{{#1}}
\definecolor{mycolour}{RGB}{0,128,0}

\title{Stabilising  viscous extensional flows using Reinforcement  Learning}

\author{Marco Vona}
\author{Eric Lauga}
\email{e.lauga@damtp.cam.ac.uk}
\affiliation{Department of Applied Mathematics and Theoretical Physics, 
University of Cambridge, Wilberforce Road, Cambridge CB3 0WA, UK}
\date{\today}
\begin{abstract}
The four-roll mill, wherein four identical cylinders undergo rotation of identical magnitude but alternate signs, was originally proposed by GI Taylor  to create local extensional flows and study their ability to deform small liquid drops. Since an extensional flow has an unstable eigendirection, a  drop located at the flow stagnation point  will have a tendency to escape. This unstable dynamics can however be stabilised  using, e.g.,    a modulation of the rotation rates of  the cylinders. Here we use Reinforcement  Learning, a branch of Machine Learning devoted to the optimal selection of actions based on cumulative rewards, in order to devise a stabilization algorithm for the four-roll mill flow.  The flow is modelled as the linear superposition of   four two-dimensional rotlets and the drop is treated as a rigid  spherical particle   smaller than all other length scales in the problem. 
Unlike previous attempts to devise control, we  take a probabilistic approach whereby speed adjustments are drawn from a probability density function whose shape is improved over time via a form of gradient ascent know as Actor-Critic method.  With enough training, our algorithm is able to precisely control the drop and keep it close to the stagnation point for as long as needed. We   explore  the impact of the physical and learning parameters on the effectiveness of the control and demonstrate the robustness of the algorithm against thermal noise. We finally show that Reinforcement  Learning can provide a control algorithm effective  for all initial positions and that can be adapted to  limit   the magnitude of the flow extension near the position of the drop.

\end{abstract}
\maketitle

\section{Introduction}\label{Introduction}

In his landmark 1934 paper, and one of his most cited works, GI Taylor proposed a device to study drop deformation and breakup in a  two-dimensional flow~\cite{taylor1934formation}. Now called the four-roll mill~\cite{higdon1993kinematics}, the apparatus  used electrical motors to rotate four identical cylinders immersed in a viscous fluid. Spinning all cylinders at the same speed (in magnitude), with adjacent cylinders rotating in opposite directions, led to an approximately extensional flow with a stagnation point at the centre. {Taylor aimed to study how the extension rate of the  flow  deformed the drop, when this was placed and kept stable at the stagnation point}.

The stagnation point at the centre of an extensional flow is a  saddle. The unstable nature of the stagnation point made the drop in 
Taylor's experiment  difficult to control, and the speed had to be varied in real time to compensate for the drop moving off in the wrong direction~\cite{taylor1934formation}. Over  fifty years later, a more systematic method to stabilize the  motion of the drop was proposed by Bentley and Leal~\cite{bentley1986experimental}.  
Using a camera to measure the drift of the drop and computer activated stepping motors to adjust the revolution rates, they modulated the rotation speeds of the cylinders using a simple feedback model where the position of the  stagnation point  was the control variable. 

This newly discovered control scheme allowed further advances in our understanding of capillary flows~\cite{eggers1997nonlinear}, in particular the deformation of drops in shear flows \cite{rallison1984deformation}. In more recent work, microfluidic devices  have been used to control the deformation of small drops
~\cite{stone2004engineering,SquireQuake_MicrofluidReview2005}. In particular,  microfluidic implementations of the four-roll mill~\cite{hudson2004microfluidic,lee2007microfluidic} and of related Stokes traps~\cite{tanyeri2010hydrodynamic,tanyeri2011microfluidic,shenoy2016stokes}
have led to pioneering techniques for trapping and manipulating on small scales.

\begin{figure}[t]
\includegraphics[width=0.45\textwidth]{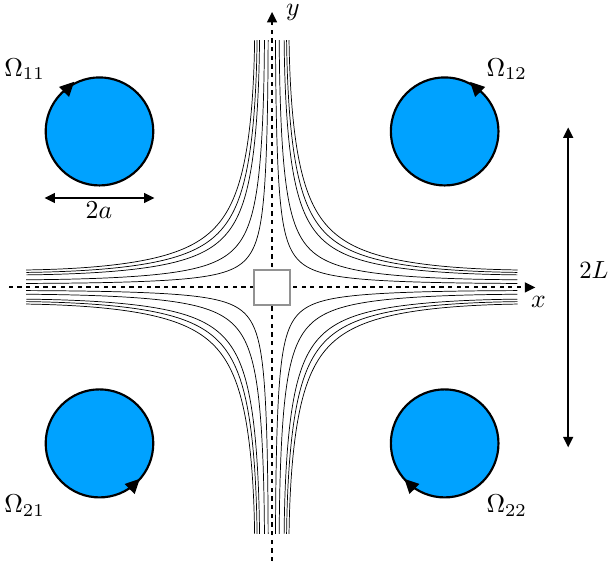}
\caption{Schematic representation of the four-roll mill setup where four rotating cylinders of radius $a$ and located on a square of side length $2L$ create an extensional flow {{near the centre of the ($x,y$) coordinate system}}. The angular velocities of the cylinders are denoted by $\Omega_{ij}$, with $i,j=1,2$. The goal of our  Reinforcement  Learning algorithm is to stabilize the motion of viscous drops inside a small square area shown schematically in grey, following the experiments in~Ref.~\cite{bentley1986experimental} (not to scale). 
The streamlines shown are from  the model flow from Eq.~\eqref{Flow Equation} in the case where the angular velocities follow the symmetric values from Eq.~\eqref{eq:eq}. 
}
\label{fig:Diagram}
\end{figure}

It is a simple mathematical exercise to show that an extension flow is unstable. Approximating the small drop by a point particle, its trajectory  in the steady flow field $\boldsymbol {u}(\boldsymbol{x})$ is solution to $\dot{\boldsymbol x}=\boldsymbol {u}(\boldsymbol{x})$. If the 
centre of the apparatus is used as the origin of the coordinate system $\boldsymbol 0$, we have  $\boldsymbol u(\boldsymbol 0)=\boldsymbol 0$ by symmetry and can therefore approximate $\dot{\boldsymbol x}\approx\nabla\boldsymbol {u}(\boldsymbol 0)\cdot\boldsymbol x$ near the origin. The extensional flow is irrotational and, since the flow is incompressible, the tensor $\nabla\boldsymbol {u}$ is symmetric and traceless. Therefore, it has real eigenvalues of identical magnitude but opposite sign,  $\lambda >0$, and $-\lambda < 0$. The centre of the extension flow is thus a saddle point, with basin of attraction parallel to the unidirectional compression and instability in all other directions (see streamlines in Fig.~\ref{fig:Diagram}).

In this paper, we aim to  design a different type of algorithm to drive the drop back when it drifts in the unstable direction. 
The natural way to correct the trajectory is to adjust the angular rotation rates of the cylinders, but to do so deterministically requires a control model describing the response of the drop  to changes in the  flow field and  for that  control scheme to be implemented in real time. This was the rationale for the   algorithm proposed in Ref.~\cite{bentley1986computer} using the position of the stagnation point as the control variable.  In this paper,  instead of a physics-based control scheme we   devise a stabilization algorithm using the framework of Reinforcement Learning~\cite{sutton2018reinforcement}.

Reinforcement Learning  is a branch of Machine Learning that allows a software agent to behave optimally in a given environment (\textit{state space}) via observation of environmental feedback. In essence, the agent explores the environment by taking \textit{actions} (which can be anything from moves in chess to steering in a self-driving car) and receiving positive or negative feedback  accordingly. Feedback comes in the form of \textit{rewards}, which, when suitably added together, make up the \textit{return} associated with the overall performance. The goal of Reinforcement  Learning is, in general, to learn how to maximize this return by improving the agent's behaviour~\cite{sutton2018reinforcement}. The learning algorithms designed to achieve this vary significantly depending on the nature of the state space 
(e.g.~continuous or discrete, finite or infinite) and on the {{agent's}} knowledge of the effect of actions. When only finitely many actions are available, finding the best behaviour is often entirely algorithmic. If however there is a continuum of states and actions, exploration is typically harder and  local improvements to the behaviour have to be found via gradient methods.  

Reinforcement Learning has found countless applications in recent years, with  significant impact already in fluid dynamics~\cite{brenner2019perspective,brunton2020machine}. For applications in  flow physics at high Reynolds number, Reinforcement Learning has been used for bulk flow control~\cite{gueniat2016statistical,rabault2019artificial}, the control of free surfaces~\cite{xie2021sloshing}   and liquid films~\cite{belus2019exploiting}, 
shape optimization \cite{viquerat2021direct}, turbulence modelling~\cite{novati2021automating}
 and sensor placement~\cite{paris2021robust}. Biological and bio-inspired applications at high high Reynolds numbers include 
  control and energy optimization  in fish swimming 
\cite{gazzola2016learning,novati2017synchronisation,verma2018efficient}, gliding and perching~\cite{novati2019controlled}
and locomotion in potential flows~\cite{jiao2021learning}. A landmark study even demonstrated how to exploit Reinforcement Learning in experimental conditions for turbulent drag reduction in flow past 
 bluff bodies~\cite{fan2020reinforcement}. 
Applications in the absence of inertia have been motivated by biological problems in navigation and locomotion, and include 
optimal navigation and escape of self-propelled swimmers \cite{colabrese2017flow,gustavsson2017finding,colabrese2018smart}, learning to swim 
~\cite{tsang2020self,liu2021mechanical} and to perform chemotaxis~\cite{hartl2021microswimmers} or even active  cloaking \cite{mirzakhanloo_esmaeilzadeh_alam_2020}. Reinforcement Learning was also incorporated in experiments using artificial microswimmers navigating in noisy environments~\cite{muinos2021reinforcement}.

 In our study, we show how to use the framework of Reinforcement Learning  to successfully control the position of a drop in a model of the four-roll mill setup.   The flow is modelled as the linear superposition of  four two-dimensional rotlets and the drop treated as a rigid  spherical particle  smaller than all other length scales in the problem.  Our state space is a small neighbourhood of the unstable equilibrium in the resulting two-dimensional extension flow, and our actions  consist of varying the speed of the cylinders at each time step. We  reward actions depending on whether the speed adjustment moves us towards the origin during the time step. Since this is a low-Reynolds-number setup, we can assume that the flow and the drop both respond instantaneously to speed modulation, so that the outcome of an action depends only on the drop's current position, and not on its current speed or acceleration. The  chosen learning algorithm is a classic Actor-Critic method based on gradient ascent. Actions are determined by a set of parameters that are varied, at every time step,  in the direction of an estimate of the gradient of performance with respect to these  parameters
 
After introducing the flow model in \S\ref{sec:setup}, we give a quick overview of Reinforcement  Learning in \S\ref{sec:RL} along with {{a}} description of our algorithm. The various physical and learning parameters are summarised in \S\ref{sec:parameters}. We then demonstrate  in \S\ref{Algorithm in Action} that, with the right  choice of parameters, our algorithm is effective at stabilising the drop from any initial drift. Next, in \S\ref{sec:results}, we explore the impact of the various physical and learning parameters on the effectiveness of the algorithm. Finally, in \S\ref{Robustness} we address the robustness of the algorithm against thermal noise, its  ability to provide a global policy for all initial positions, and how to modify the algorithm to enable control of the magnitude of the flow extension near the position of the drop.

\section{Flow model and trajectories}\label{sec:setup}
\subsection{Flow}
As a prototypical  device generating  an extension flow, we consider a simple model for a two-dimensional four-roll mill. The  flow is generated by four identical cylinders centred at the corners of a square of side length $2L$. All lengths  are non-dimensionalised by $L$ so that the centres of the cylinders are located at $(\pm 1, \pm 1)$ in a $(x,y)$ Cartesian coordinate system (see Fig.~\ref{fig:Diagram}). Motivated by application in microfluidics, we assume that the rotation rates of the cylinders are small enough that all inertial effects in the fluid can be neglected. We further assume that the cylinders are sufficiently long and  far away from each other that we can approximate   the flow created by each cylinder as a two-dimensional rotlet~\cite{Batchelor1970,kimbook}, i.e.~by the exact solution for the two-dimensional Stokes flow outside an isolated cylinder in an infinite fluid. {{The flow induced by each cylinder at position $\boldsymbol x$ is hence given by}} 
\begin{equation}\label{eq:totalflow}
\boldsymbol u(\boldsymbol x)= a^2\Omega_{ij}\boldsymbol e_z\times\frac{\boldsymbol x-\boldsymbol c_{ij}}{|\boldsymbol x-\boldsymbol c_{ij}|^2},
\end{equation}
where $a$ is the dimensionless radius of the cylinder and where $\Omega_{ij}$ and $\boldsymbol c_{ij}$ are, respectively, the angular velocity and the location of the centre of the cylinder; note that $\Omega_{ij}>0$ indicates anticlockwise rotation (see Fig.~\ref{fig:Diagram}). In the limit where $a\ll 1$, we  may approximate the flow near the centre of the device as a linear superposition of the four  flows from each cylinder, so that for $|\boldsymbol x|$ small,
\begin{equation}\label{Flow Equation}
\boldsymbol u(\boldsymbol x)= a^2\sum_{i,j}\Omega_{ij}\boldsymbol e_z\times\frac{\boldsymbol x-\boldsymbol c_{ij}}{|\boldsymbol x-\boldsymbol c_{ij}|^2}.
\end{equation}
Note that this two-dimensional flow is irrotational.

As in Taylor's original,  the case where
\begin{equation}\label{eq:eq}
\Omega_{11}=-\Omega_{12}=-\Omega_{21}=\Omega_{22},
\end{equation}
  leads to a purely extensional flow near the origin, since the off-diagonal entries of the velocity gradient $\nabla \boldsymbol u$ are $0$ by symmetry. Our Reinforcement  Learning algorithm will then modify the individual angular velocities $\Omega_{ij}$   independently in order to correct  trajectories (see \S\ref{sec:results}), so Eq.~\eqref{eq:eq}  holds only before speed control is applied. 
  
  \subsection{Drop motion}\label{sec:dropmotion}

  We  model the viscous drop, transported by the flow and for which we want to achieve stable motion, as a rigid spherical particle of radius $r$ (we thus assume that the drop is very rigid and the Capillary number small enough to not deform it significantly). Its centre, located at $\boldsymbol x(t)$, evolves in time according to Fax\'en's law~\cite{kimbook}
 \begin{equation}\label{dropode}
\frac{\d \boldsymbol x}{\d t}=\left(1+\frac{r^2}{6}\nabla^2\right)\boldsymbol u(\boldsymbol x),
\end{equation}
where the flow $\boldsymbol u$ is given by Eq.~\eqref{eq:totalflow}. Note that for this choice of flow the Fax\'en term $\propto\nabla^2\boldsymbol u$ is identically zero because the flow is both incompressible and irrotational and thus $\nabla^2\boldsymbol u=\nabla(\nabla\cdot\boldsymbol u)-\nabla\times(\nabla\times\boldsymbol u)=\boldsymbol 0$. In the absence of noise,  we integrate Eq.~\eqref{dropode} numerically with the Runge-Kutta RK4 method. 
In \S\ref{sec:noise} we 
 also incorporate thermal noise (i.e.~Brownian motion) as relevant to the dynamics of   small drops.

\section{Reinforcement  Learning algorithm}\label{sec:RL}
\subsection{Fundamentals of Reinforcement  Learning}\label{Fundaments of RL}

We begin  by introducing some  terminology that underpins the rest of the work; the reader is {{referred}} to the classical book by Sutton and Barto for a detailed treatment~\cite{sutton2018reinforcement}. In Reinforcement  Learning, agents take actions that depend on their current state, and get rewarded accordingly. The mathematical basis is that of  Markov Decision Processes~\cite{bellman1957markovian}, which consist of the following:

(1) A \textit{state space} $S$ to be explored, with realization $s$.

(2) An \textit{action space} $A$ (or $A_s$, since it may vary between states), with realization $a$, which comprises the moves available at each state.

(3) A probability density function (or mass function, if $S$ is countable) $P(s'|s,a):S\times S\times A\to \mathbb R$, which determines the probability of transitioning from state $s$ to state $s'$ after taking action $a$. This probability never changes during the process.

(4) A \textit{reward function} $R(s',s,a):S\times S\times A\to\mathbb R$, which gives the reward earned  after transitioning from $s$ to $s'$ {{through}} action $a$.

The actions are drawn from a p.d.f. (or mass function, if the action sets $A_s$ are countable) $\pi(a|s):A\times S\to \mathbb R$ know as the \textit{policy}. This is the function that determines behaviour.
Exploration takes place in discrete time steps. At time step $t$, the agent lands in state $s_t$ and takes action $a_t$, which takes it to state $s_{t+1}$ according to the distribution $P$. 
The probability of landing in a given state is a function of the current state and the choice of action, so transitions have the Markov Property. {{If the probability distribution $P$ or the reward function $R$ are not known to the agent,}} this is referred to as \textit{model-free} Reinforcement  Learning.

Since we want the agent  to behave in a specific way, we introduce a notion of \textit{return} {$G_t$ from time step $t+1$ onwards, given that we are starting from state $s_t=s$ at time $t$}. We define $G_t=\sum_{k=1}^{\infty}\gamma^{k-1}R_{t+k}$, where {{$R_t$ is the reward earned at time $t$ and}} $\gamma\in [0,1]$ is known as \textit{discount factor}. Multiplication by $\gamma^{k-1}$ ensures convergence if rewards are well behaved and captures the uncertainty associated with long-term rewards.

From $G_t$ we can define the {{\textit{state value function}}} $v_{\pi}(s)=\mathbb E_{\pi}[G_t|s_t=s]$, which is the expected return starting from state $s$ and following $\pi$ (we thus use  $\mathbb E$ to denote expected values in what follows). Our goal is to find (or at least  to approximate) the policy $\pi_*$ which maximizes $v_{\pi}$, i.e.~the choice of actions leading to maximum return.

 \subsection{Choice of Markov Decision Process}\label{Choice of MDP}
We now    describe the simplest version of the algorithm used in this study, with some improvements summarised in \S\ref{Further Improvements}. 
In the specific viscous flow problem considered here, we wish  to learn how to modify the motion of the cylinders in order to manoeuvre the drop towards the origin from a \textit{fixed} starting point $\boldsymbol x_0$.  In other words, given the default angular velocities in Eq.~\eqref{eq:eq}, an initial position $\boldsymbol x_0$ for the drop and a sequence of time steps $t_0$, $t_1$, $t_2$,..., we want to learn how to change the angular velocity vector $\boldsymbol \Omega=[\Omega_{11},\Omega_{12},\Omega_{21},\Omega_{22}]$ at each step so as to bring the drop as close to the origin as possible. We will discuss how to extend this strategy to \textit{all} initial positions later in the paper.

We start by  assuming that, at each time step, the angular velocity vector $\boldsymbol \Omega$ changes instantaneously and that the drop's position can be computed exactly and with no delay. To make speed adjustments without the use of Reinforcement  Learning, we would need to know how a given change in angular velocity affects the trajectory before changing the speed, which is computationally unfeasible. Using Reinforcement  Learning, in contrast, we can limit ourselves to observing how the drop reacts to a speed change in a given position and learn through trial and error.

We can now formulate the problem in terms of a Markov Decision Process, following points $1-4$ in \S\ref{Fundaments of RL}: 

(1)  We choose {the \textit{state space} $S$  to be a square of dimensionless side length $0.1$ centred at the origin (shown schematically in Fig.~\ref{fig:Diagram}). The drop starts somewhere inside this square and needs to reach the origin while moving inside this square. If the drop ever leaves this region during a run, we terminate execution because the drop has wandered too far. The exact size of the region can be changed depending on the accuracy needed, and it will come into play   when we try to find a general strategy that does not work just for $\boldsymbol x_0$.}

(2) {The \textit{action space} associated with state $s$ consists of all allowed changes to $\boldsymbol \Omega$ in that particular state. Since we are going to use a gradient ascent method to determine  the optimal changes, it is important to  keep the action space as small as possible. If, for example, we allowed ourselves to act on all four cylinders at every time step {our policy would become a function of position and range over all  speed adjustments. Such complexity would be hard to approximate, especially with a probabilistic gradient method}. Instead, {{with}} our algorithm, we only  act on one cylinder at a time, thereby reducing the dimension of $A$ to $1$. We split the plane into four quadrants (one per cylinder) and whenever the drop is located  in a specific quadrant we only allow  the cylinder ahead of it in the clockwise direction to modify its rotation speed  (this is illustrated graphically in  the insets of Fig.~\ref{fig:illustrate}B, C).  An action consists of changing the angular velocity for that specific cylinder, $\Omega_{ij}$, to some other value in a prescribed interval $[\Omega_{ij}-w,\Omega_{ij}+w]$, where $w>0$ gives the size of the ``wiggle'' room and is chosen in advance for all cylinders (see below for more details). At the end of the time step, we instantaneously reset the velocity of this cylinder, so that the effect of the subsequent action only depends on the final position of the drop. Since we have no inertia, transitions obey   the Markov Property.}

(3) {With regard to the probability density function $P$, {{in the absence of thermal noise each position}} $\boldsymbol x_{t+1}$ is a deterministic function of  $\boldsymbol x_t$ and of the action $a_t$ and is independent of {{$\boldsymbol x_{\tau}$ and $a_{\tau}$ for all $\tau<t$}}. So we can write  $\boldsymbol x_{t+1}=F(\boldsymbol x_t,a_t)$ for some $F$, and hence the probability density function is a delta function, i.e.~$P(\boldsymbol x'_{t+1}|\boldsymbol x_t,a_t)=\delta[\boldsymbol x'_{t+1}-F(\boldsymbol x_t,a_t)]$. Note that, if we knew $F$ exactly, we would also know how actions affect the trajectory, so the problem would be trivially solved. The reason why some sort of control algorithm is needed  is precisely that $F$ cannot be easily determined.} It is worth mentioning that, had we included inertia in the problem,
{{we would have needed to add the drop's velocity and acceleration to the state space in order for $P$ to be well-defined;}} 
this increase in dimensionality would have made the problem harder.

(4) {For the rewards, we need to favour actions that move the particle closer to the origin, and punish ones that bring it further away from it. We thus choose to reward each speed adjustments in relation to the the drop's subsequent displacement vector. Our choice of \textit{reward function} is given by
\begin{equation}\label{Reward Function}
R(\boldsymbol x_{t+1},\boldsymbol x_{t},a_t)=\exp\left\{-p\left[1+\frac{(\boldsymbol x_{t+1}-\boldsymbol x_{t})\cdot(\boldsymbol x_{t})}{||\boldsymbol x_{t+1}-\boldsymbol x_{t}||\cdot ||\boldsymbol x_{t}||}\right]\right\},
\end{equation}
where $p>0$ is a dimensionless parameter designed to tune the peakedness of the   function  inside the exponential; we explore below how the performance depends on the value of $p$ (the value {{$p=1$}} will be chosen for most results).  To aid intuition, note that the reward function can also be  written as $R=\exp\{-p[1-\cos\theta_t]\}$, where $\theta_t$ is the angle that the displacement vector makes with the inward radial vector $-\boldsymbol x_t$. The reward is thus maximal when $\theta_t=0$ (inward radial motion) and minimal when $\theta_t=\pi$ (outward radial motion). 
 We found it  important that our reward function evaluate  actions on a continuous scale. If, for example, we were to assign a value of $1$ to moves that point us within some angle of the right direction and $0$ to everything else, the algorithm would  regard all bad moves as equally undesirable and have difficulty learning. 
{An exponential dependence was chosen over other options, such as a piecewise linear function, in order to reduce the number of free parameters.
}} 

 \subsection{Choice of algorithm}

For our Reinforcement  Learning algorithm, we choose a  classic Actor-Critic method based on gradient ascent~\cite{sutton2018reinforcement}. The  ``Actor'' refers to the policy, {{which encodes behaviour}}, while the ``Critic'' refers to the value function, which measures expected returns.  We introduce parametric approximations of  both the policy and the state value function, and then, at each time step, vary the parameters in the direction of an estimate of the gradient of performance with respect to the  parameters.

\subsection{Actor part of algorithm}

 We wish to determine the optimal policy for this problem, i.e.~the  p.d.f.~$\pi(a|\boldsymbol x)$ that maximizes $v_{\pi}(\boldsymbol x_{0})$ {{for some fixed $\boldsymbol x_0$}}. We introduce a parametric policy of the form $\hat\pi(a|\boldsymbol x;\boldsymbol C)$, 
 where $\boldsymbol C$ is some array that characterizes the policy, and we then use gradient ascent on $\boldsymbol C$ to find a local optimum for $v_{\hat\pi}(\boldsymbol x_{0})$ (in all that follows, when we use a subscript $\hat\pi$  in the value functions, it will always indicate implicitly a  dependence on $\C$).  In other words, if we define $J(\boldsymbol C)=v_{\hat\pi}(\boldsymbol x_{0})$,  we will seek to optimize for $\boldsymbol C$ by iterating $\boldsymbol C_{t+1}= \boldsymbol C_t+\alpha_t\nabla J(\boldsymbol C)|_{\C_t}$. 
 This will allow us to  improve the policy at every time step (so-called \textit{online learning}). This is referred to as the ``Actor'' part of the algorithm, because the policy generates behaviour.

 Computing the gradient $\nabla J$ may appear difficult a priori, but can be achieved using  a powerful   result  known as the policy gradient theorem, proven in Ref.~\cite{sutton2018reinforcement} for countable action spaces. This theorem states that at time $t$ the gradient is equal to
\begin{equation}\label{PGThm}
\nabla J(\boldsymbol C_t)=\mathbb E_{\hat\pi}[
A_{\hat\pi}(\boldsymbol x_t,a_t)\nabla_{\boldsymbol C}\log\hat\pi(a_t|\boldsymbol x_t;\boldsymbol C)|_{\C_t}]
\end{equation}
where 
\begin{equation}
A_{\hat\pi}(\boldsymbol x_t,a_t)= Q_{\hat\pi}(\boldsymbol x_t,a_t)-v_{\hat \pi}(\boldsymbol x_t)
\end{equation}
 is known as the \textit{advantage function} and we have introduced
\begin{equation}
Q_{\hat\pi}(\boldsymbol x_t,a_t)=\mathbb E_{\hat\pi}[G_t|\boldsymbol x_t,a_t]
\end{equation}
 which is  the \textit{action-state value function}.

The result in Eq.~\eqref{PGThm} suggests a practical way to implement an algorithm to determine the parameters of the optimal policy. Specifically, we drop the expectation and, after drawing $a_t$ from the current policy (more on this below),  iterate on the parameters of the policy as
 \begin{equation}\label{eq:algoC}
\boldsymbol C_{t+1}=\boldsymbol C_t+\alpha_t A_{\hat\pi}(\boldsymbol x_t,a_t)\nabla_{\boldsymbol C}\log\hat\pi(a_t|\boldsymbol x_t;\boldsymbol C)|_{\C_t}.
\end{equation} 
{{at each time step}}. Note that this leads to an \textit{unbiased} estimate of the policy gradient because the expected value of the update is the true value of the gradient. 
 
 \subsection{Parametric policy}

{{We now need to write down an expression for $\hat\pi$, i.e. our guess for the true optimal policy.}}
{
 Since we have a logarithm in Eq.~\eqref{PGThm}, it is convenient to write $\hat\pi$ in the form 
 \begin{equation}\label{eq:policyhat}
\hat\pi(a|\boldsymbol x;\boldsymbol C)=\frac{1}{K}\exp[f(\x,a; \boldsymbol C)],
\end{equation}
 where $\boldsymbol x=(x,y)$ and where} 
 {
  \begin{equation}
 K(\x;\C)=\int_A  \exp[f(\x,a; \boldsymbol C)] \mathrm d a.
\end{equation}
}
 {
For fixed ($\boldsymbol x$,$\boldsymbol C$), this {{ensures that $\hat\pi(a|\boldsymbol x, \C)$}} is a p.d.f.~for $a$. Here $f$ can be any convenient function, and in what follows  we take it to be a polynomial in the parameters. Specifically, we take $\boldsymbol C$ to be an $n\times m\times p$ array and set 
\begin{equation}\label{fapprox}
f(\x,a; \boldsymbol C)=\sum_{i,j,k}  C_{ijk}x^{i-1}y^{j-1}a^{k-1}.
\end{equation}
As noted before, this can only work if the action space is not too large. If, for example, we could act on multiple cylinders simultaneously, we would need a more complex Ansatz for $f$ as well as a higher-dimensional array $\C$, which would make gradient ascent   harder.
Then, at time step $t$, the \textit{score function} $\nabla_{\boldsymbol C}\log\hat\pi(a_t|\boldsymbol x_t; \boldsymbol C)|_{\C_t}$ becomes a $n\times m\times p$ array $\boldsymbol T_t$ such that} 
{
\begin{align}
& T_{t,ijk}\nonumber\\
&= x_t^{i-1}y_t^{j-1}a_t^{k-1}-\frac{1}{K}\left\{\int_A\nabla_{\C}\exp[f(\x_t,a; \boldsymbol C)] \mathrm d a\right\}_{t,ijk}\nonumber\\
&=x_t^{i-1}y_t^{j-1}\left\{a_t^{k-1}-\frac{1}{K}\int_A\exp[f(\x_t,a; \boldsymbol C_t)]a^{k-1} \mathrm d a\right\}\nonumber\\
&=x_t^{i-1}y_t^{j-1}\left(a_t^{k-1}-\mathbb E_{\hat \pi}[a^{k-1}|\x_t]\right). 
\end{align} 
In practice, we generate a second action $\tilde a_t$ at time $t$ and take
\begin{equation}\label{T approximation}
T_{t,ijk}=x_t^{i-1}y_t^{j-1}\left(a_t^{k-1}-\tilde a_t^{k-1}\right),
\end{equation}}
where we use the subscript $t$  to indicate that this is its value at time step $t$. Then our  algorithmic update for $\boldsymbol C$ in Eq.~\eqref{eq:algoC} becomes 
\begin{equation}\label{eq:alog2}
\C_{t+1}= \boldsymbol C_t+\alpha_t A_{\hat\pi}(\boldsymbol x_t,a_t)\boldsymbol T_t.
\end{equation}
Note that other choices for $f$ are of course possible, a truncated Fourier series being the obvious one, but we found that Eq.~\eqref{fapprox} was computationally faster. Note also that $R_t+\gamma v_{\hat\pi}(\boldsymbol x_{t+1})-v_{\hat\pi}(\boldsymbol x_t)$ is an unbiased estimate of the advantage function $A_{\hat\pi}$, since 
\begin{eqnarray}
\mathbb E_{\hat\pi}[R_t+\gamma v_{\hat\pi}(\boldsymbol x_{t+1})|\boldsymbol x_t,a_t]-v_{\hat\pi}(\boldsymbol x_t)\nonumber\\
=Q_{\hat\pi}(\boldsymbol x_t,a_t)-v_{\hat\pi}(\boldsymbol x_t)
=A_{\hat\pi}(\boldsymbol x_t,a_t).
\end{eqnarray}
 Therefore we can replace $A_{\hat\pi}$ in Eq.~\eqref{eq:alog2} and iterate  
\begin{equation}\label{CUpdate}
\boldsymbol C_{t+1}= \boldsymbol C_t+\alpha_t [R_t+\gamma v_{\hat\pi}(\boldsymbol x_{t+1})-v_{\hat\pi}(\boldsymbol x_t)]\boldsymbol T_t.
\end{equation}

\subsection{Critic Part of Algorithm}
The second, or ``Critic'', part of the algorithm deals with the approximation of the value function. 
To make use of Eq.~\eqref{CUpdate}, we replace the state value function for our policy $v_{\hat\pi}$ with another parametric approximation $\hat v_{\hat\pi}(\boldsymbol x;\boldsymbol D)$, where $\boldsymbol D$ is once again an array. The goal is then to determine $\boldsymbol D$ which minimizes the distance $H(\boldsymbol D)=\mathbb E_{\hat\pi}[(v_{\hat\pi}-\hat v_{\hat\pi})^2]$. This can be done numerically by using gradient descent with the update rule 
\begin{equation}
\boldsymbol D_{t+1}= \boldsymbol D_t-\frac{1}{2}\beta_t\nabla H(\boldsymbol D)|_{\D_t},
\end{equation}
 where $\beta_t>0$. Assuming that we can take the gradient inside the expectation, we have
\begin{equation}
-\frac{1}{2} \nabla H(\boldsymbol D)|_{\D_t}= \mathbb E_{\hat\pi}\left[(v_{\hat\pi}-\hat v_{\hat\pi})\nabla_{\boldsymbol D} \hat v_{\hat\pi}(\boldsymbol x_t;\boldsymbol D)|_{\D_t}\right].
\end{equation}
After  replacing the expectation with the corresponding unbiased estimate, we then obtain the the gradient algorithmic update rule
 \begin{equation}
\boldsymbol D_{t+1}=\boldsymbol D_t+\beta_t(v_{\hat\pi}-\hat v_{\hat\pi})\nabla_{\boldsymbol D} \hat v_{\hat\pi}(\boldsymbol x_t;\boldsymbol D)|_{\D_t}.
 \end{equation}
  
 Finally, since $v_{\hat\pi}(\boldsymbol x_t)=R_t+\gamma v_{\hat\pi}(\boldsymbol x_{t+1})$, we can use the approximation  $v_{\hat\pi}(\boldsymbol x_t)=R_t+\gamma \hat v_{\hat\pi}(\boldsymbol x_{t+1}; \boldsymbol D_t)$ to get the final form of the update rule as
 \begin{eqnarray}
\boldsymbol D_{t+1}=\boldsymbol D_t+ \beta_t\delta_t\nabla_{\boldsymbol D} \hat v_{\hat\pi}(\boldsymbol x_t;\boldsymbol D)|_{\D_t}.
 \end{eqnarray} 
 where
 \begin{equation}
 \delta_t=R_{t}+\gamma \hat v_{\hat\pi}(\boldsymbol x_{t+1};\boldsymbol D_t)-\hat v_{\hat\pi}(\boldsymbol x_t;\boldsymbol D_t).
 \end{equation}
 
  Similarly to Eq.~\eqref{fapprox}, we take $\boldsymbol D$   to be an $r\times s$ array {{and}}
\begin{equation}
\hat v_{\hat\pi}(\boldsymbol x_t,\boldsymbol D)=\sum_{i,j} D_{ij}x_t^{i-1}y_t^{j-1}.
\end{equation} 
Then $\nabla_{\boldsymbol D} \hat v_{\hat\pi}(\boldsymbol x_t;\boldsymbol D)|_{\D_t}=\boldsymbol Q_t$ with $ Q_{t,ij}=x_t^{i-1}y_t^{j-1}$, and the update becomes 
\begin{equation}
\boldsymbol D_{t+1}=\boldsymbol D_t+\beta_t\delta_t\boldsymbol Q_t.
\end{equation}
To ensure convergence, it is customary to make $\alpha_t$ and $\beta_t$ decay geometrically, which we do here by setting $\alpha_t=\alpha \gamma^t$ and $\beta_t=\beta \gamma^t$ where $\alpha$ and $\beta$ are constants and where $\gamma$ is the discount factor~\cite{sutton2018reinforcement}.

\subsection{Summary of algorithm}\label{Summary}
To summarise, the  algorithm we implement works as follows:
\begin{enumerate}
\item{Choose the step size constants $\alpha$ and $\beta$;}
\item{{Initialise the arrays $\boldsymbol C_0$ and $\boldsymbol D_0$ to $\0$ and choose the drop position  $\boldsymbol x_0$ at $t=0$.}}
\item{At time step $t$, draw a random action from $\hat\pi(a_t|\boldsymbol x_t; \boldsymbol C_t)$ and record the corresponding reward {{$R_t$}} and next state $\boldsymbol x_{t+1}$.  }
\item{Update the two parameters as
\begin{subeqnarray}\label{eq:summaryupdate}
\boldsymbol C_{t+1}=\boldsymbol C_t+\alpha\gamma^t\delta_t\boldsymbol T_t\\\boldsymbol D_{t+1}=\boldsymbol D_t+\beta\gamma^t\delta_t\boldsymbol Q_t\end{subeqnarray}
where $\delta_t=R_{t}+\gamma \hat v_{\hat\pi}(\boldsymbol x_{t+1};\boldsymbol D_t)-\hat v_{\hat\pi}(\boldsymbol x_t;\boldsymbol D_t)$.}
\item{Repeat until convergence.}
\end{enumerate}

\subsection{Sampling from $\hat\pi$}\label{Rejection Sampling}
The final problem we need to address is how to sample from  Eq.~\eqref{eq:policyhat}, i.e.~randomly chose an action from the approximate policy,  without {{knowing the normalising function $K(\x,\C)$}}. For this we can use a technique know as \textit{rejection sampling}. Say we want to sample {{from}} a   p.d.f.~$p(x)$ on $I\subseteq\mathbb R$ but we only know $A p(x)$ for an unknown $A>0$. We then consider {{a second (known)}} p.d.f. $q(x)$ on $I$ and we take $B>0$ so that $B q(x)\geq A p(x)$ for all $x$ in $I$. We generate $X\sim q$ and compute 
\begin{equation}\label{RSEquation}
\alpha=\frac{A  p(X)}{B  q(X)} 
\end{equation}
Then we generate $Y\sim U([0,1])$ and if $Y\leq \alpha$ we accept $X$, otherwise we reject it. Then, conditional on being accepted, we have $X\sim p$. A proof of this algorithm is given   in Appendix~\ref{sec:proof}.

In our case we can take $q$ to be the uniform distribution; to generate $A$ it then suffices to find an upper bound for $\frac{1}{K} \exp(f)$, which is straightforward if we have a bound on each of $x$, $y$, $a$. The only drawback of this method is that if $\alpha$ is very small it may take a long time to find an acceptable $a$. {To get around this, we generate $a_t\in[-L,L]$ and then take the speed adjustment to be $ {wa_t}/{L}$, where $w>0$ is the wiggle room size and $L$ is the half-width of the state space. By taking $a_t=\mathcal O(L)$, we can easily find a reasonably small upper bound for $f$, which helps to keep $\alpha$ relatively large.
 }

 \subsection{Time delays and noise}\label{Further Improvements}
Two aspects were finally added to the algorithm in order to make it {{physically realistic}}.  First, we dropped the mathematical assumption that the cylidnders can change speed instantaneously. Instead, we assumed that the computer takes a  time $t_{\rm lag}$ to determine the position of the drop, accelerates the cylinder over a time ${t_1}$ and resets the velocity to its initial value over a time ${t_2}$. All these delays are included in the same time step. In all cases, we assume that cylinders  speed up or slow down with constant angular acceleration. Secondly, we allowed the drop to no longer follow a deterministic trajectory by adding  thermal noise, as explained in \S\ref{sec:dropmotion}. This allows us to test the application of the algorithm to setups on small length scales.  
 
  \section{Physical and Reinforcement  Learning parameters}
\label{sec:parameters} \subsection{Episodes and batches}

In order to estimate the best policy for a given starting point $\boldsymbol x_0$, we would need to apply the Actor-Critic method outlined above to an infinitely long sequence of states starting from $\boldsymbol x_0$. In practice, however, this sequence has to terminate, so it is customary to instead run a number of \textit{episodes} (i.e.~sequences of states) of fixed length $t_{\text{max}}$ from $\boldsymbol x_0$ and  to apply the Actor-Critic algorithm to every state transition as usual. In each new episode, we then  use the latest estimates of $\boldsymbol C$ and $\boldsymbol D$ as parameters.

 A straightforward way to assess the speed at which learning is done is to group episodes into \textit{batches} (typically of $100$) and examine the effectiveness of the values of $\boldsymbol C$ and $\boldsymbol D$ given by the learning process at the end of each batch. To do this, we use the values of $\boldsymbol C$ and $\boldsymbol D$ so obtained to run a separate series of $100$ episodes starting from $\boldsymbol x_0$ during which no learning occurs. We then compute the average final distance to origin, $\overline{|\boldsymbol {x}_{\text{f}}|}$, as a proxy for the effectiveness of the control algorithm in  bringing {{the drop back to the origin}}.  Note that if we were estimating the policy for a real-world experiment, we would stop running batches as soon as $\overline{|\boldsymbol {x}_{\text{f}}|}$ becomes suitably small, and then use the resulting values of  the parameters as our practical control algorithm. 

\subsection{Physical parameters }\label{sec:physicalp}
In order to run the algorithm we need to fix values for the   parameters describing the flow and the physical setup in which the drops moves. In order to use a classical study  as benchmark, we take these physical parameters from Bentley and Leal's four-roll mill control study~\cite{bentley1986experimental}. {{This leads to the   choices of parameters as follows (all dimensions below will thus refer to their paper):}}

\begin{enumerate}
\item 
 Length scales are non-dimensionalised by half the   distance between the centre of the cylinders  ($6.35\text{ cm}$) and time scales by the angular velocities (which we take to be $\Omega = \pm 0.125\text{ s}^{-1}$) from Ref.~\cite{bentley1986experimental}.

\item{The radius $a$ of the cylinders was $5.08 \text{ cm}$ in Ref.~\cite{bentley1986experimental} so the non-dimensionalised cylinder radius is $a=0.8$}. {Note that in our theoretical approach for the flow in Eq.~\eqref{Flow Equation}, we assumed $a\ll 1$. The experiments  in Ref.~\cite{bentley1986experimental} are therefore at the limit of what can be captured by the simple hydrodynamic theory, which is  however useful in what follows to demonstrate a proof of concept of our control approach.}
\item{The angular velocity vector $\boldsymbol \Omega$  defined in \S\ref{Choice of MDP} and satisfying Eq.~\eqref{eq:eq} is now given by $[-1,1,1,-1]$.}
\item{For the  initial position $\boldsymbol x_0$ of the drop, we {{start by taking}} $\boldsymbol x_0=[-0.03,0.02]$ to test the functionality of our algorithm, and we then extend to different starting points in \S\ref{Algorithm in Action}. }
\item{Four time {parameters} have to be set: the  times step ($dt$), the lag $(t_{\rm lag})$ and the  delays   (${t_1}$ and  ${t_2}$). In Ref.~\cite{bentley1986experimental} the lag was about $0.1\text{ s}$, and the motors took {{roughly}} $0.05\text{ s}$ to modulate speed. We will thus take  non-dimensionalised values {{$t_{\rm lag}=0.0125$, $t_1=t_2=0.005$, $dt=0.025$}}. Note that because the algorithm is only rewarded for moving the drop in the right direction,  the exact values of the time parameters actually do not matter, as long as they all get scaled by the same factor. In other words, we still expect the algorithm to learn if we replace the four time scales $dt$, $t_{\rm lag}$, $t_1$ and $t_2$ with $\lambda  dt$, $\lambda t_{\rm lag}$, $\lambda  t_1$ and $\lambda  t_2$ for some $\lambda>0$. Reducing the time parameters gives of course more control over the drop, since we adjust its trajectory more frequently.
}
\item{We need to set the dimensionless rotation wiggle rooms, i.e.~the range of angular velocities we allow for the cylinders. For simplicity these are same for all cylinders, set to be $\pm 0.7$ unless specified otherwise;  we will study how varying this parameter affects performance in \S\ref{sec:varyw}.}
\end{enumerate}
The flow velocity is given by Eq.~\eqref{Flow Equation} and the trajectory of the drop is obtained by integrating Eq.~\eqref{dropode} with the   speed parameters during each step. For numerical integration, we employed the fourth order Runge-Kutta scheme RK4 with step size $dt/20$.

\subsection{Actor-Critic parameters}
The Reinforcement  Learning algorithm outlined above contains also a number of parameters: 
\begin{enumerate}
\item{The side length of the state space, which we always take to be $0.1$ in dimensionless units (see \S\ref{Choice of MDP}).}
\item{The discount factor $\gamma$, which appears in the state-value function. Computationally, this comes into play in both the policy and the state value function update (see \S\ref{Summary}).} We will study the impact of varying the value of $\gamma$ in \S\ref{sec:varygamma}.
\item{The peakedness $p$  of the reward function in Eq.~\eqref{Reward Function}. We will study the impact of varying  $p$ in \S\ref{sec:varyp}.}
\item{The step size constants  $\alpha$ and $\beta$ used in the update part of the algorithm, Eq.~\eqref{eq:summaryupdate}. For simplicity, we   assume $\alpha=\beta$ and will study the impact of varying their values   in \S\ref{sec:varyalpha}.}
\item{The size of $\boldsymbol C$ and $\boldsymbol D$. For simplicity, we take $\boldsymbol C$ to be $N\times N$ and $\boldsymbol D$ to be $N\times N\times N$, and choosing the value of $N$ is discussed in \S\ref{sec:varyN}.}
\item{The length $t_{\text{max}}$ of each episode; this will be varied in \S\ref{sec:varyt}.}
\end{enumerate}

In our exploration, we start by choosing the values of the algorithm parameters  randomly and then we vary them one at a time to see how they affect accuracy and learning speed. We use the mean final distance to the origin, $\overline{|\boldsymbol {x}_{\text{f}}|}$, to monitor the algorithm's success  in bringing back the drop to the centre of the flow. When we compute this quantity, all parameters remain the same as in the training episodes, and $\boldsymbol C$ and $\boldsymbol D$ are held fixed.
 When find a local optimum for one parameter, we keep it fixed at that value in subsequent simulations, thereby leading to a set of parameters which  should optimize performance, at least locally. This  exploration of parameters will be further discussed in \S\ref{sec:results}.

\section{Illustration of learned  policy} 
\label{Algorithm in Action}
 
In this first section of results, we demonstrate the effectiveness of Reinforcement  Learning in stabilising the motion of the drop when lags and delays are included (but not thermal noise). We first let the algorithm practice with a given starting point and then   simulate a trajectory to assess performance
(i.e.~the practical control of the drop's motion).  We will  illustrate the details of the   learning process in \S\ref{sec:results} and the robustness of the algorithm in \S\ref{Robustness}. The codes used as part of this study have also been posted on GitHub~\cite{GH} where they are freely available.

{{We assume here that all physical parameters are as in \S \ref{sec:physicalp} and take $t_{\text{max}}=40$}}. The  parameters of the Reinforcement Learning algorithm, which will be examined in detail in \S\ref{sec:results}, are taken to be  $\alpha=\beta=10$, $\gamma=0.95$, $p=1$, $N=4$. We also set the rotation wiggle room to be $70\%$ of the default angular velocity.

  \begin{figure*}[t]
\includegraphics[width=0.75\textwidth]{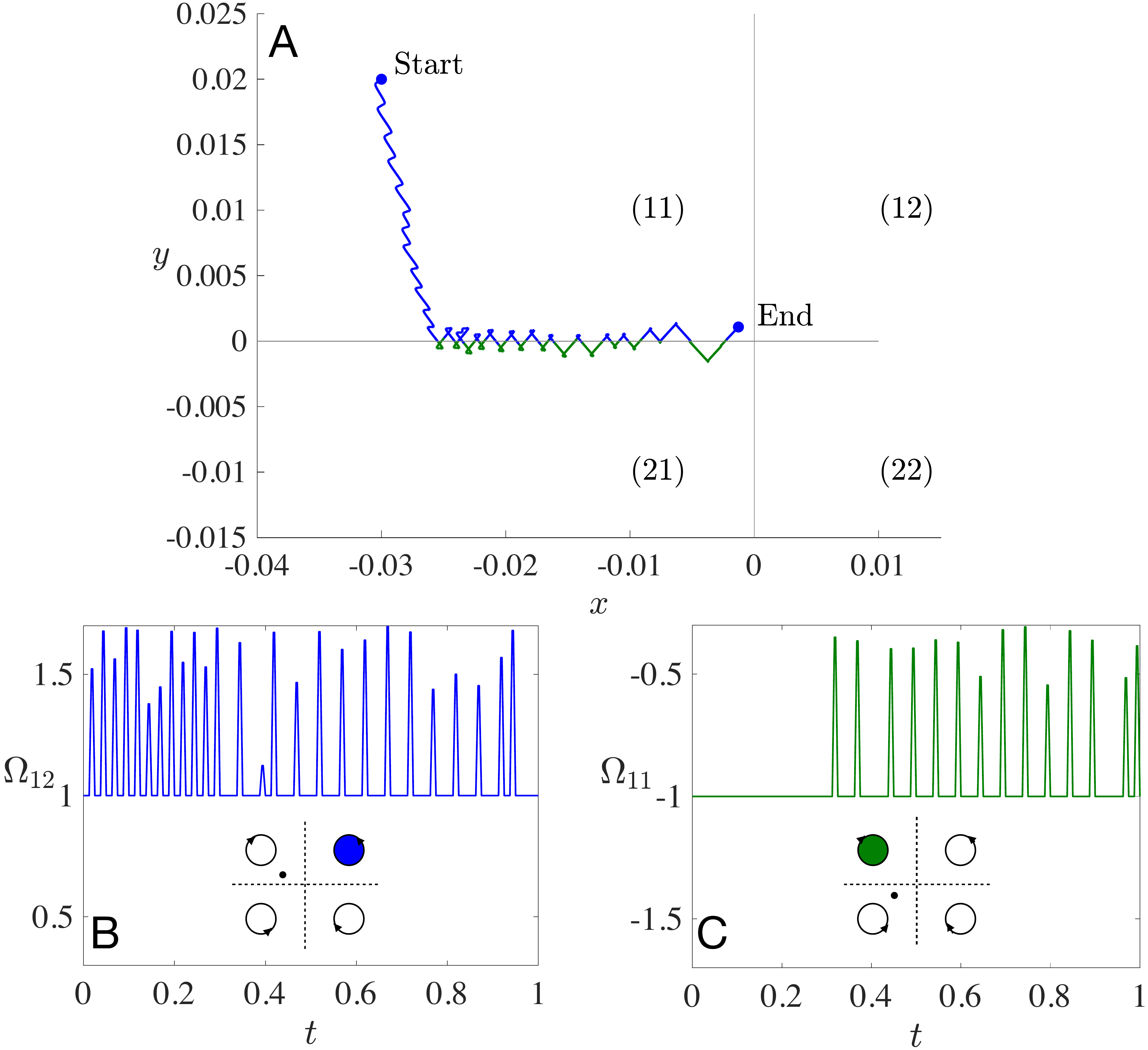}
\caption{Illustration of the learned policy for a drop starting at the dimensionless location $\boldsymbol x_0=[-0.03,0.02]$ with the algorithmic choices $\alpha=\beta=10$, $\gamma=0.95$, $p=1$, $N=4$ and for  rotation wiggle rooms of $70\%$ of the default angular velocity of each cylinder. 
A: Drop undergoing a zig-zag motion, first toward the $x$ axis and then toward the origin. The blue portion of the trajectory indicates when the control is done by changing the value of $\Omega_{12}$ (its value is plotted in B, with rotating cylinder shown as inset) while in the green curve the change is done by  tuning the value of  $\Omega_{11}$ (its value is shown in C, with rotating cylinder shown as inset).}
\label{fig:illustrate}
\end{figure*}

We start the drop  at the dimensionless location $\boldsymbol x_0=[-0.03,0.02]$,    estimate $\boldsymbol C$, $\boldsymbol D$ over $700$ {episodes} and then use the learned policy to plot the trajectory of the controlled drop motion. {{In an  experiment, one would use the algorithm to estimate the policy and then apply the control policy until the drop is sufficiently close to the origin, after which the cylinder could resume spinning at their default velocities}}.  Results are shown in Fig.~\ref{fig:illustrate}A, with 
a  movie of the   motion available in Supplementary Material~\cite{SM}. Since the drop starts in the $(11)$ quadrant (see Fig.~\ref{fig:Diagram}), the motion is initially only affected by  $\Omega_{12}$, i.e.~the rotation {{rate}} of the cylinder  ahead of it in the clockwise direction.  {{We show in Fig.~\ref{fig:illustrate}B and Fig.~\ref{fig:illustrate}C  the time-evolution of $\Omega_{12}$ and $\Omega_{11}$ in  blue and green, respectively}}. The use of the green and blue  in the trajectory from Fig.~\ref{fig:illustrate}A highlights the parts of the  trajectory where each  cylinder undergoes a change in its rotation speed (the corresponding cylinder is indicated in the  insets of Fig.~\ref{fig:illustrate}B, C). {{The final distance from the origin was about $0.0010$, which is smaller that the non-dimensionalised value of $0.0078$ required in the experiments of Ref.~\cite{bentley1986computer}}}.

{{The results in Fig.~\ref{fig:illustrate} suggest a simple physical interpretation of the policy}}. {{In the absence of control, the drop would be advected towards $x\to -\infty$, $y\to 0$ from its initial position}}  (see streamlines  shown in Fig.~\ref{fig:Diagram}). The policy obtained via  the Reinforcement Learning algorithm causes the angular speed $\Omega_{12}$ to undergo bursts of small increases above its steady value (typically 50\% in magnitude); when $\Omega_{12}$ is increased, the drop is seen to undergo a small diagonal displacement towards the $x$ axis, while when $\Omega_{12} = 1$ the drop experiences a small amount of free motion. By alternating  between the  two, the drop is eventually able to reach the  $x$ axis. Note that the  sharp corners in some of the pathlines are a consequence of the absence of inertia. After reaching the $x$ axis, the drop crosses into the $(21)$ quadrant, {{where $\Omega_{11}$ (the only cylinder we can now act on) undergoes similar small bursts in order to bring the drop back to the $x$ axis}}. {{The net result of the alternating actions of $\Omega_{11}$ and $\Omega_{12}$ is a zig-zag motion on both sides of the $x$ axis, which eventually brings the drop acceptably close to the origin}}.
Note that when taking the non-dimensionalization into account, {{the motion displayed in Fig.~\ref{fig:illustrate} would take about $8$ s}} in the original experimental setup of Ref.~\cite{bentley1986computer}.

  \begin{figure}[t]
\includegraphics[width=0.45\textwidth]{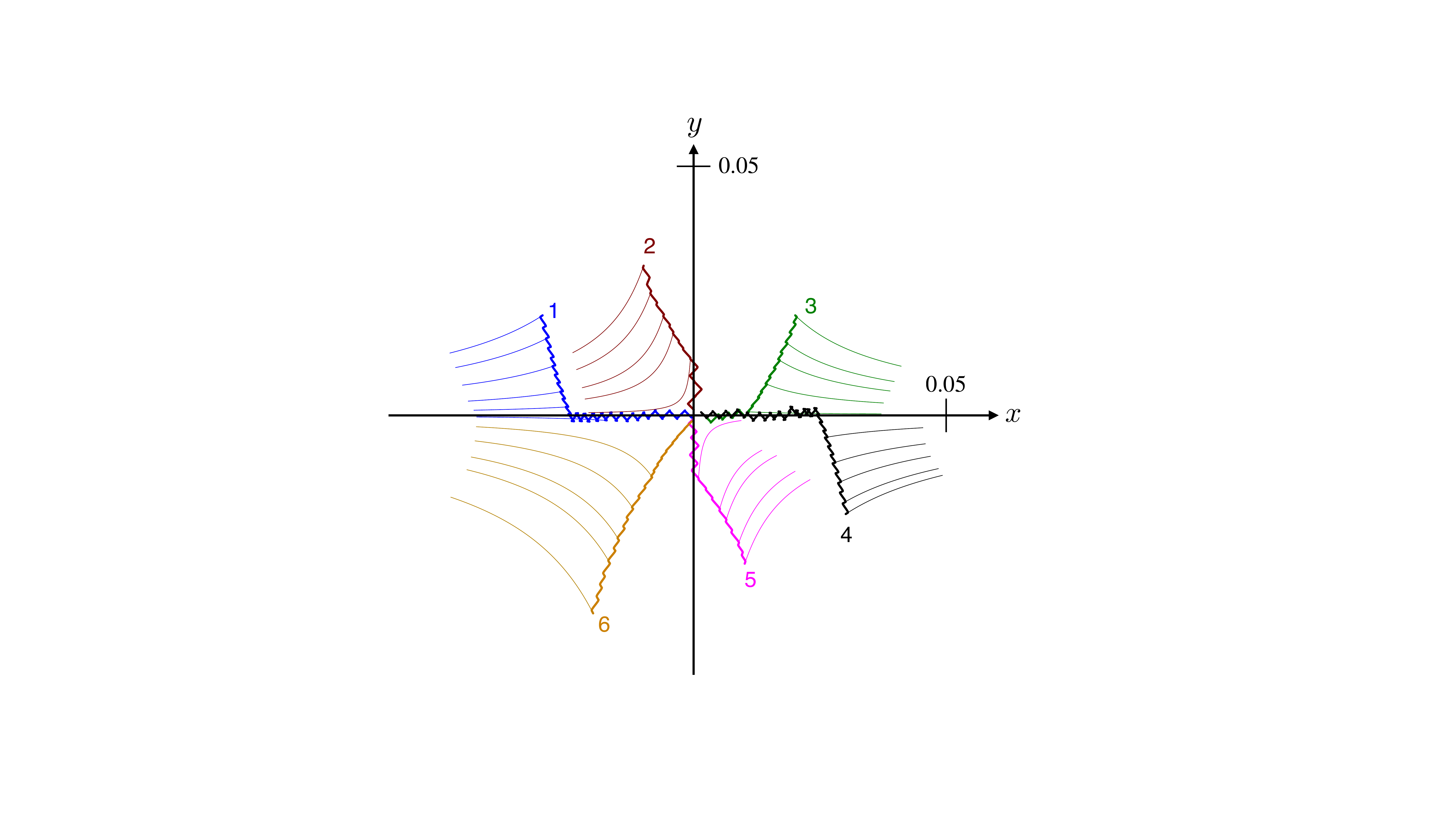}
\caption{Illustration of  different policies  learned  after starting from six different locations: $\boldsymbol x_1=[-0.03,0.02]$, $\boldsymbol x_2=[-0.01,0.03]$, $\boldsymbol x_3=[0.02,0.02]$, $\boldsymbol x_4=[0.03,-0.02]$, $\boldsymbol x_5=[0.01,-0.03]$, $\boldsymbol x_6=[-0.02,-0.04]$. The learning parameters are $\alpha=\beta=10$, $N=4$, $\gamma=0.95$ and $p=1$. In each case we use as many batches as required to get the final position of the drop below  $0.0015$. We plot the resulting policy in thick {{solid}}, while {{the}} thin lines show what the drop would do {{in the absence of speed}} control.}
\label{fig:SixTrajectories}
\end{figure}

We next illustrate how the algorithm performs from different  starting points, {{as well as}} how trajectories change depending on the initial position. We again take $\alpha=\beta=10$, $N=4$, $\gamma=0.95$, $p=1$ and choose the same time parameters as before. We consider six different starting points $\boldsymbol x_k$ located in the four quadrants, specifically $\boldsymbol x_1=[-0.03,0.02]$, $\boldsymbol x_2=[-0.01,0.03]$, $\boldsymbol x_3=[0.02,0.02]$, $\boldsymbol x_4=[0.03,-0.02]$, $\boldsymbol x_5=[0.01,-0.03]$, $\boldsymbol x_6=[-0.02,-0.04]$. 
  The algorithm trains for each point separately, i.e.~it computes a different policy for each value of $\boldsymbol x_k$. For each starting point, we allow the algorithm to practice on as many batches as needed until {{$\overline{|\boldsymbol x_{\text{f}}|}$}} drops below $0.0015$. This never took more than $7$ batches, i.e.~$700$ episodes. 

We show in Fig.~\ref{fig:SixTrajectories} {{the trajectories resulting from the learned policies}} (to reduce crowding, each trajectory terminates as soon as the distance from the origin at the end of a time step becomes smaller than $0.002$). In each case, the controlled motion of the drop is shown in thick solid, while {{the thins lines correspond to the paths that the drop would follow if it were not for speed control (these paths  coincide with the streamlines in Fig.~\ref{fig:Diagram})}}. In all cases, we see that the algorithm succeeds in  {{bringing}} the drop back to the origin. {{All trajectories present small diagonal drifts caused by bursts of increased rotation, separated by free motion along the streamlines}}. Computationally, points further away from the origin required more training; for example, {{finding the trajectory}} starting from $\boldsymbol x_3$  in Fig.~\ref{fig:SixTrajectories} only required $200$ training runs, while the one starting from $\boldsymbol x_4$ took $700$.

We close by emphasising  that, in all cases illustrated in Fig.~\ref{fig:SixTrajectories}, the policy  is different for each starting point; in \S\ref{Robustness} we  investigate whether it is possible to find a global policy that is effective for all starting points. 

 \section{Learning process and parameters}
\label{sec:results}
{{The previous section demonstrated the effectiveness of Reinforcement Learning in controlling the motion of the drop}}. We now investigate how accuracy and learning speed depend on the various parameters used by the algorithm. Then we   examine  in \S\ref{Robustness} how the algorithm deals with noise and with finding a global policy. {As explained in \S\ref{sec:parameters}, learning is assessed by running a fixed number of batches ($n_{\text{batches}}$) of $100$ episodes and plotting the values of the average final distance $\overline{|\x_{f}|}$ in each batch.  Since results are random, we quantify the uncertainty in each learning curve $\overline{|\x_{f}|}_i$ ($1\leq i\leq n_{\text{batches}}$) by generating it twice with the same parameters and returning the average relative error
\begin{equation}\label{Average Error}
\Delta=\frac{1}{n_{\text{batches}}}\sum_{i=1}^{n_{\text{batches}}}\frac{\left|\overline{|\x_{f}|}_i^{(1)}-\overline{|\x_{f}|}_i^{(2)}\right|}{\overline{|\x_f|}_i^{(1)}}.
\end{equation}
Since the setup is four-fold symmetric, we restrict our attention to the case where the drop starts out in the second quadrant (denoted by $11$, see Fig.~\ref{fig:illustrate}).
}

\subsection{Varying the discount Factor $\gamma$}
\label{sec:varygamma}
\begin{figure}[t]
\includegraphics[width=0.45\textwidth]{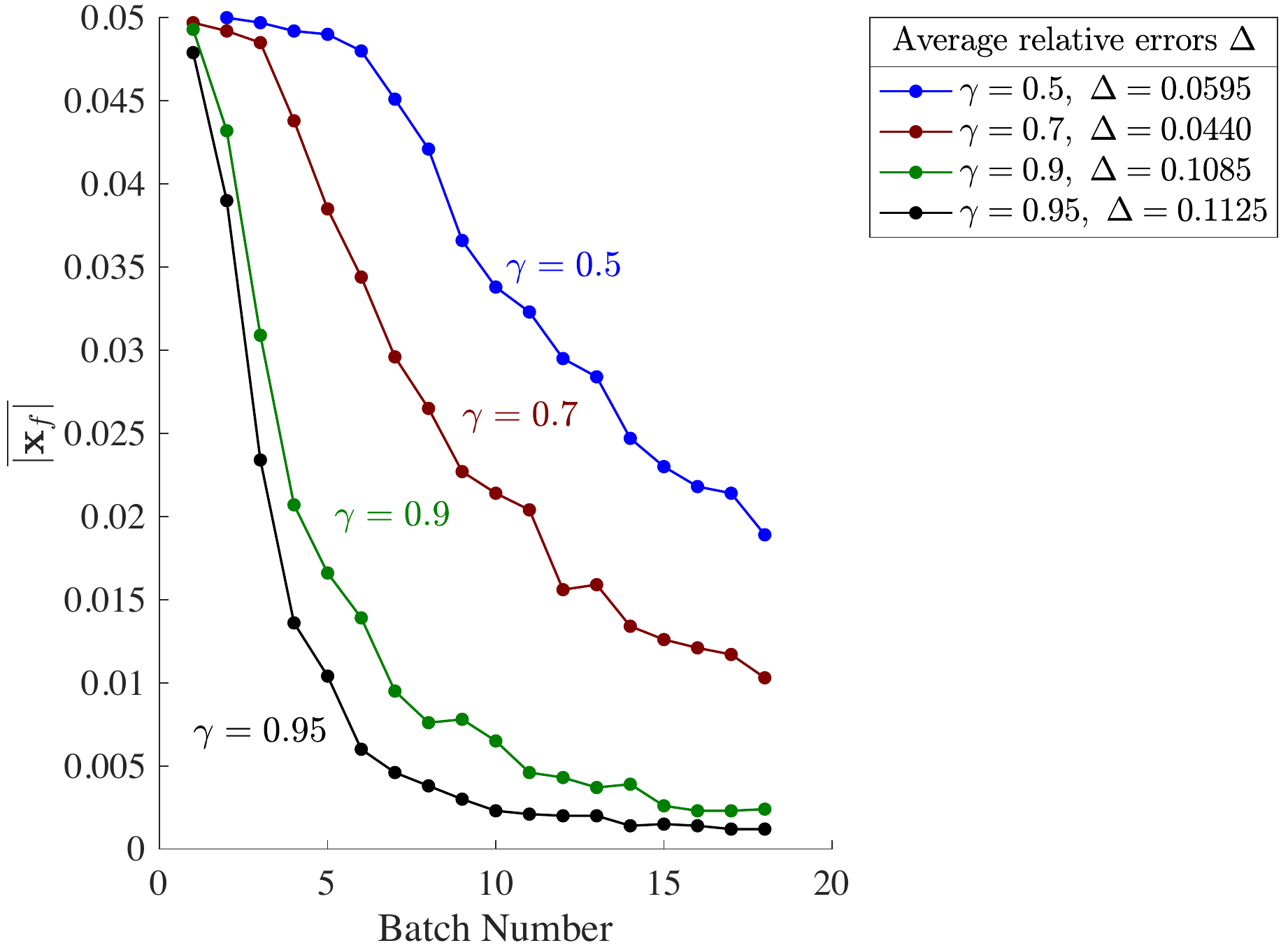}
\caption{Average final distance of the drop from the stagnation point, $\overline{|\boldsymbol {x}_{\text{f}}|}$,  as a function of the  batch number for different values of the discount factor, $\gamma$ (see text for the values of the other parameters). }
\label{fig:VariousGammas}
\end{figure}

We start by setting $p=1$, $\alpha=\beta=5$, $N=4$, $t_{\text{max}}=40$ and aim to find the value of the discount factor $\gamma$ which causes $\overline{|\boldsymbol {x}_{\text{f}}|}$ to decrease the fastest. We ignore the values $\gamma=0$ and $\gamma=1$, since
$\gamma=0$ would result in a very shallow one-step lookahead, and $\gamma=1$ would not ensure $\gamma^{t_{\text{max}}}\approx 0$, while decay is required in the updates of the Actor-Critic method.   {{In Fig.~\ref{fig:VariousGammas} we plot the average final distance $\overline{|\boldsymbol {x}_{\text{f}}|}$ as a function of the batch number for different values of $\gamma$ in the range $(0,1)$}}. Clearly  performance improves steadily with $\gamma$, showing that we can base our choice of actions on long-term predictions; the larger the value of $\gamma$ the more we penalize bad actions far ahead in the future, since the $k$th reward gets discounted by $\gamma^{k-1}$. {The average relative errors are small, indicating that the variance within each learning curve is likely to be small.} Since it gave the best performance, we take $\gamma=0.95$ in what follows.

\subsection{Varying the peakedness $p$ of the reward function}
\label{sec:varyp}
 \begin{figure}[t]
\centering{\includegraphics[width=0.45\textwidth]{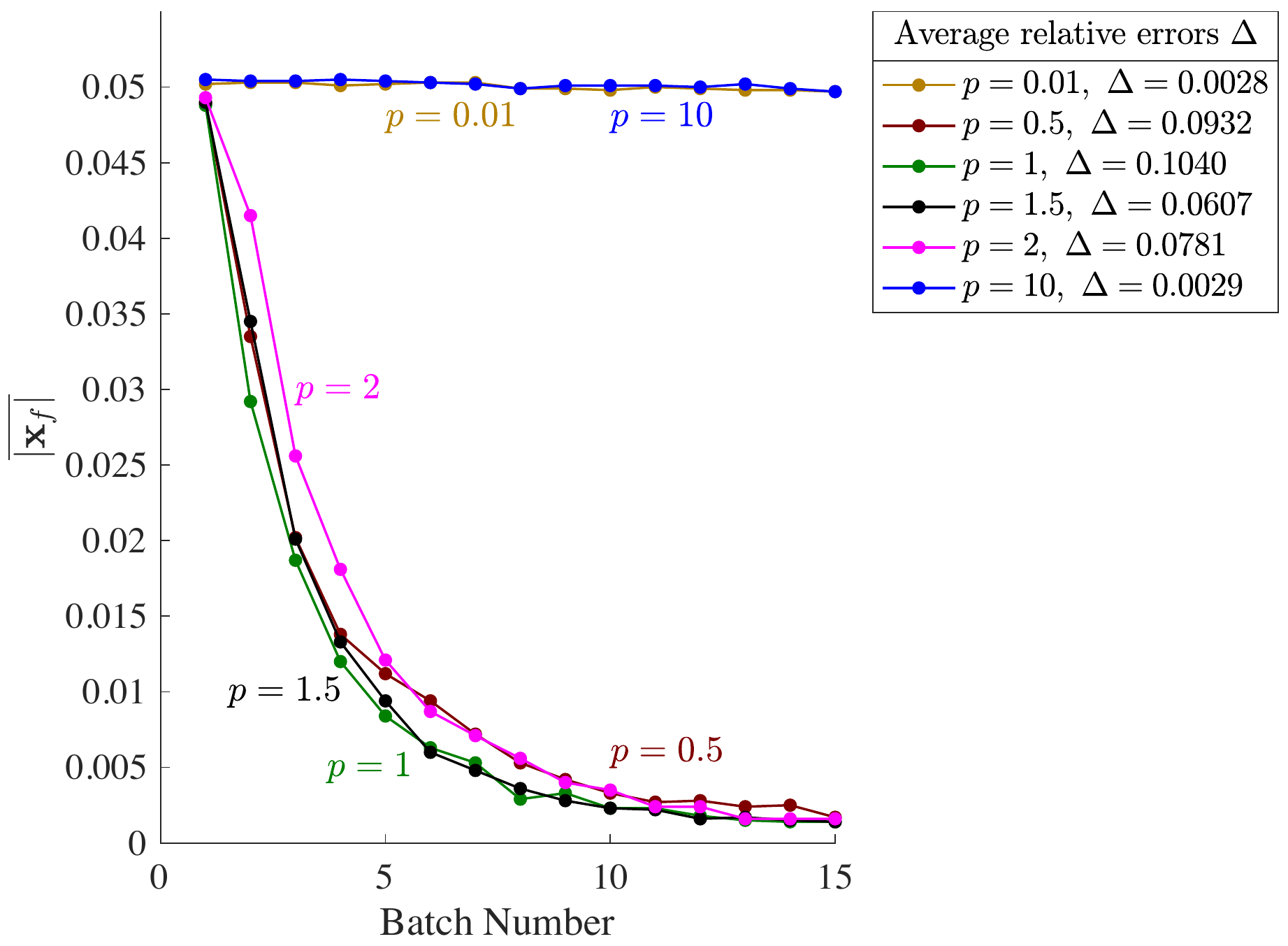}}
\caption{Average final distance of the drop from the stagnation point, $\overline{|\boldsymbol {x}_{\text{f}}|}$,  as a function of the  batch number for different values of peakedness $p$ of the reward function (see text for the values of the other parameters). }
\label{fig:VariousPs}
\end{figure}

To address the impact of the peakedness $p$ of the reward function, {{in Fig.~\ref{fig:VariousPs} we plot}} the learning curves obtained by running the algorithm with the values $\alpha=\beta=5$, $N=4$, $t_{\text{max}}=40$, $\gamma=0.95$ and   different values of $p$. Small values of $p$, such as $p=0.01$, do not adequately discriminate between actions, while very large values (e.g.~$p=10$) hinder exploration by treating all bad actions as equally undesirable, and also take longer to run. {The average relative errors are again very small, which makes us confident that the displayed curves are representative samples. Within a small window from $p=0.5$ to $p=1.5$, the learning speed increases slightly, but since results are all very similar we keep $p=1$ in what follows.}

\subsection{Varying the gradient ascent parameters $\alpha$, $\beta$}
\label{sec:varyalpha}

\begin{figure}[t]
\centering{\includegraphics[width=0.45\textwidth]{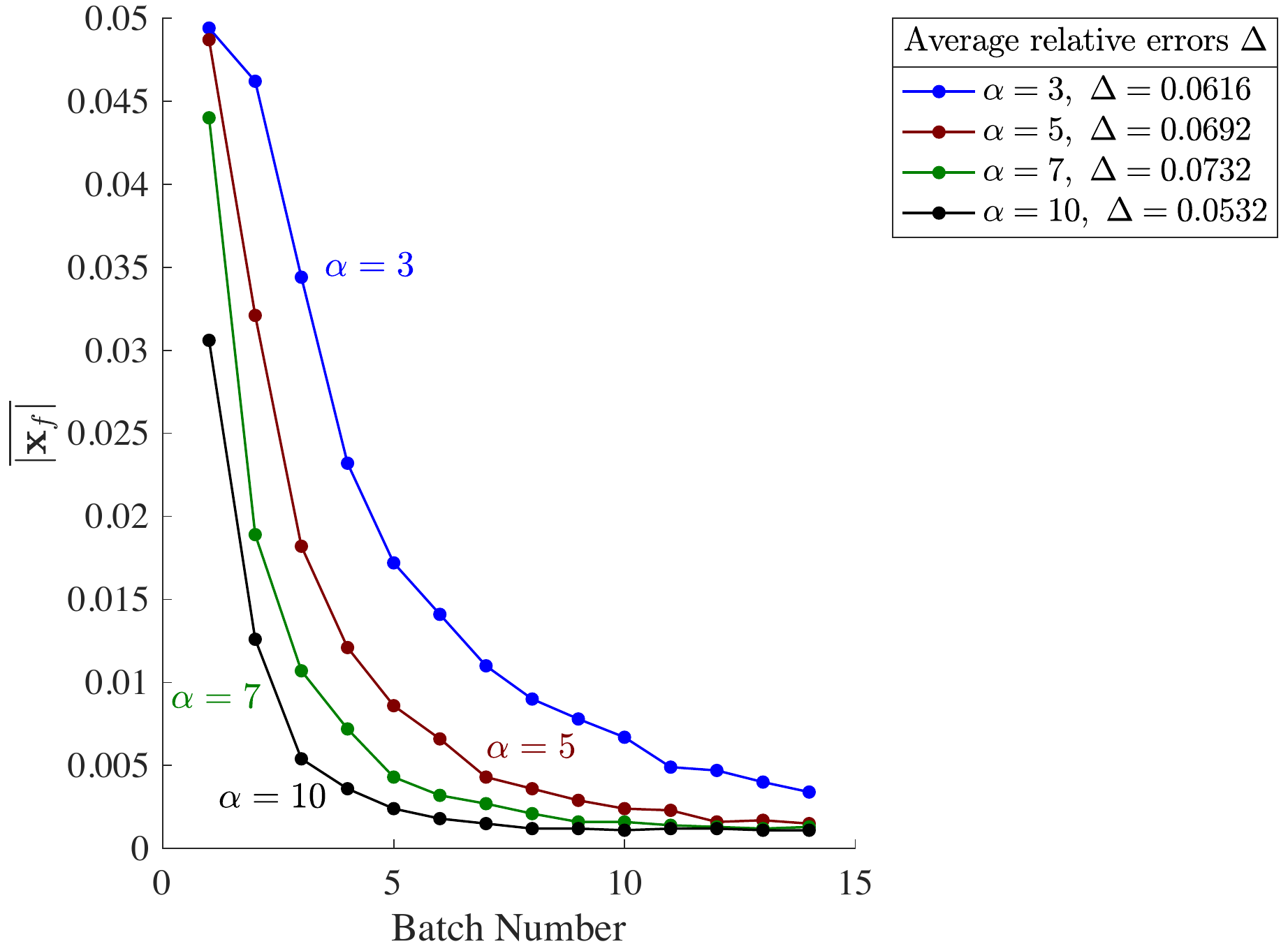}}
\caption{Average final distance of the drop from the origin, $\overline{|\boldsymbol {x}_{\text{f}}|}$,  as a function of the  batch number for different values of the  gradient ascent parameters $\alpha=\beta$ (see text for the values of the other parameters). }
\label{fig:PerformanceVsAlpha}
\end{figure}

Even with the previous choices of parameters, it still takes approximately $1000$ episodes to reach a final accuracy of {{$\overline{|\boldsymbol {x}_{\text{f}}|}=0.005$}} ($10$ batches of $100$ episodes, or more). Out of all parameters, we found that the gradient ascent parameters $\alpha$ and $\beta$ have the biggest impact on learning speeds. When chosen correctly, they can  reduce the number of training episodes  to just a few  hundred.  To demonstrate this, we take 
$t_{\text{max}}=40$, $N=4$, $p=1$, $\gamma=0.95$  {{and}} monitor the final average  {{distance}} $\overline{|\boldsymbol {x}_{\text{f}}|}$ for various values of $\alpha=\beta$; the resulting learning curves are shown in Fig.~\ref{fig:PerformanceVsAlpha}.  {{Performance}} increases steadily with $\alpha=\beta$. The values $\alpha=\beta=10$ lead to a steep learning curve, dropping below $0.01$ after only {{$3$}} iterations. The only real constraint on these  {{parameters}} is that they cannot be arbitrarily large, because  {{for}} $\gamma=0.95$ the Actor-Critic algorithm  may give very large entries for $\boldsymbol C$ and $\boldsymbol D$, {{making it hard to find a suitable $B$}} for the rejection sampling part (\S\ref{Rejection Sampling}). {Furthermore, the final gradient ascent update in each episode has size $\mathcal O(\gamma^{t_{\text{max}}}\alpha)$. If we want this to be reasonably small for $t_{\text{max}}=40$ and $\gamma=0.95$, e.g.~less than $1$, we should take $\alpha\lesssim 8$.} {As in the previous simulations, the relative errors are seen to be very small.} We therefore settle on the values $\alpha=\beta=10$ in what follows.

\subsection{Varying the size $N$ of policy and value function arrays}
\label{sec:varyN}
\begin{figure}[t]
\centering{\includegraphics[width=0.45\textwidth]{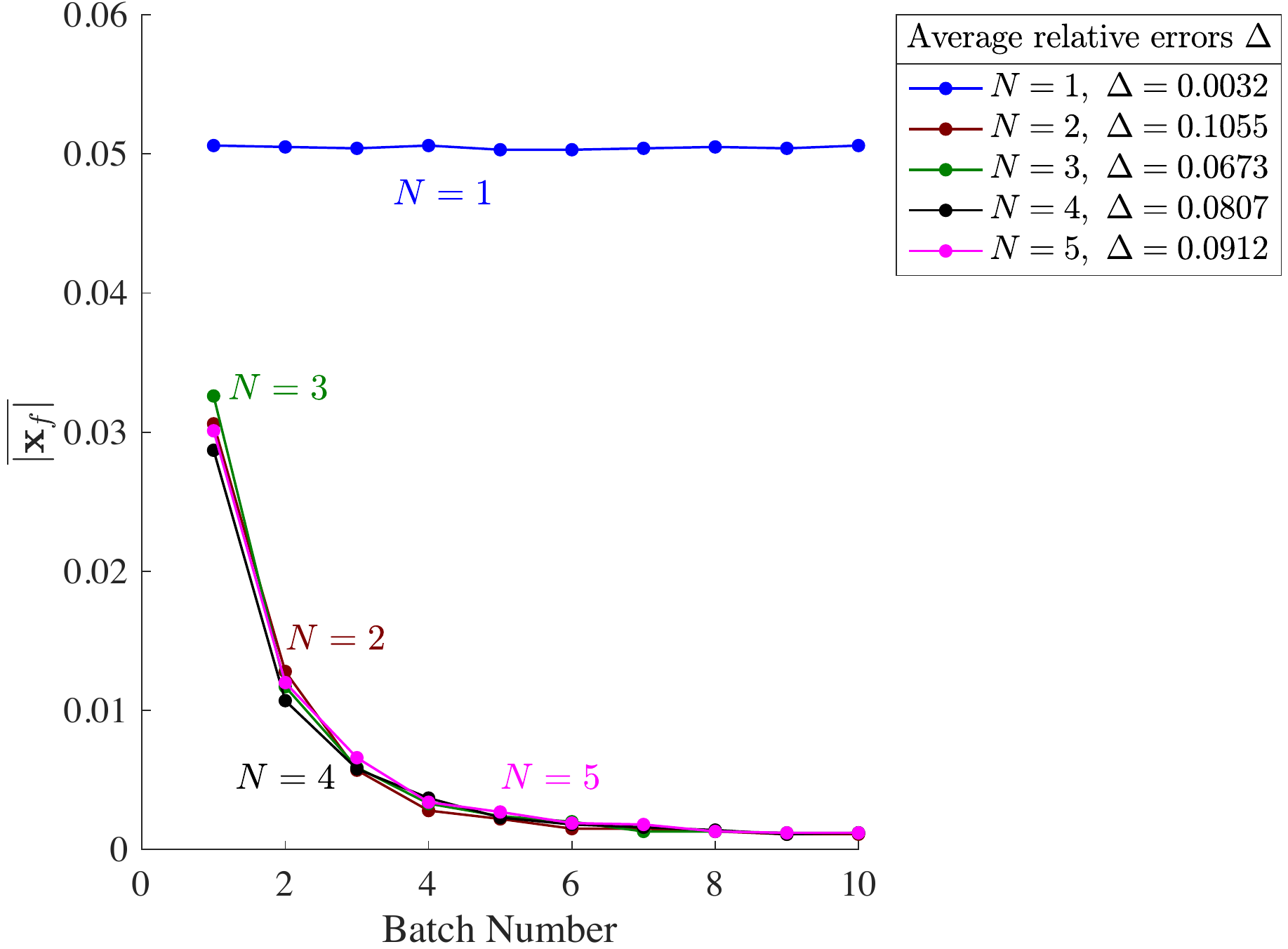}}
\caption{{Average final distance of the drop from the origin, $\overline{|\boldsymbol {x}_{\text{f}}|}$,  as a function of the batch number for different values of the array size $N$ (see text for the values of the other parameters).} }
\label{fig:PerformanceVsN}
\end{figure}

We next examine the impact of the size $N$ of the  policy and value function arrays $\C$ and $\D$. To see how this parameter affects the final accuracy and the learning speed, we choose $5$ different values of $N$ and run $10$ batches for each value {(the other parameters are kept at $t_{\text{max}}=40$, $p=1$ and $\alpha=\beta=10$, $\gamma=0.95$). The learning curves are displayed in Fig.~\ref{fig:PerformanceVsN}. We see that the choice $N=1$ performs poorly since $\hat\pi$ becomes a uniform distribution; the remaining values give very similar results, with small relative errors $\Delta$, so we keep $N=4$ in what follows. }

\subsection{Varying the step size $dt$ and the length of episodes $t_{\text{max}}$}
\label{sec:varyt}
The accuracy of the algorithm depends strongly on the step size $dt$, with larger values leading to a poorer accuracy. Furthermore, for large values of $dt$, learning may still occur but the learning curve is no longer steadily decreasing with batch number because bad actions  can take the drop further away from the target.  
Through extensive simulations, we found empirically  that  $dt$ should be chosen so that  the particle can never move  {{by a distance larger than the desired final accuracy during a time step}}. In their original paper, 
Bentley and Leal  state that a {{dimensionless}} final distance of $0.0078$ is enough for their experiments ~\citep{bentley1986computer}, {{and since $\overline{|\boldsymbol {x}_{\text{f}}|}$ was consistently below this threshold in the previous sections, our chosen $dt$ is sufficiently small}}.

\begin{figure}[t]
\centering{\includegraphics[width=0.45\textwidth]{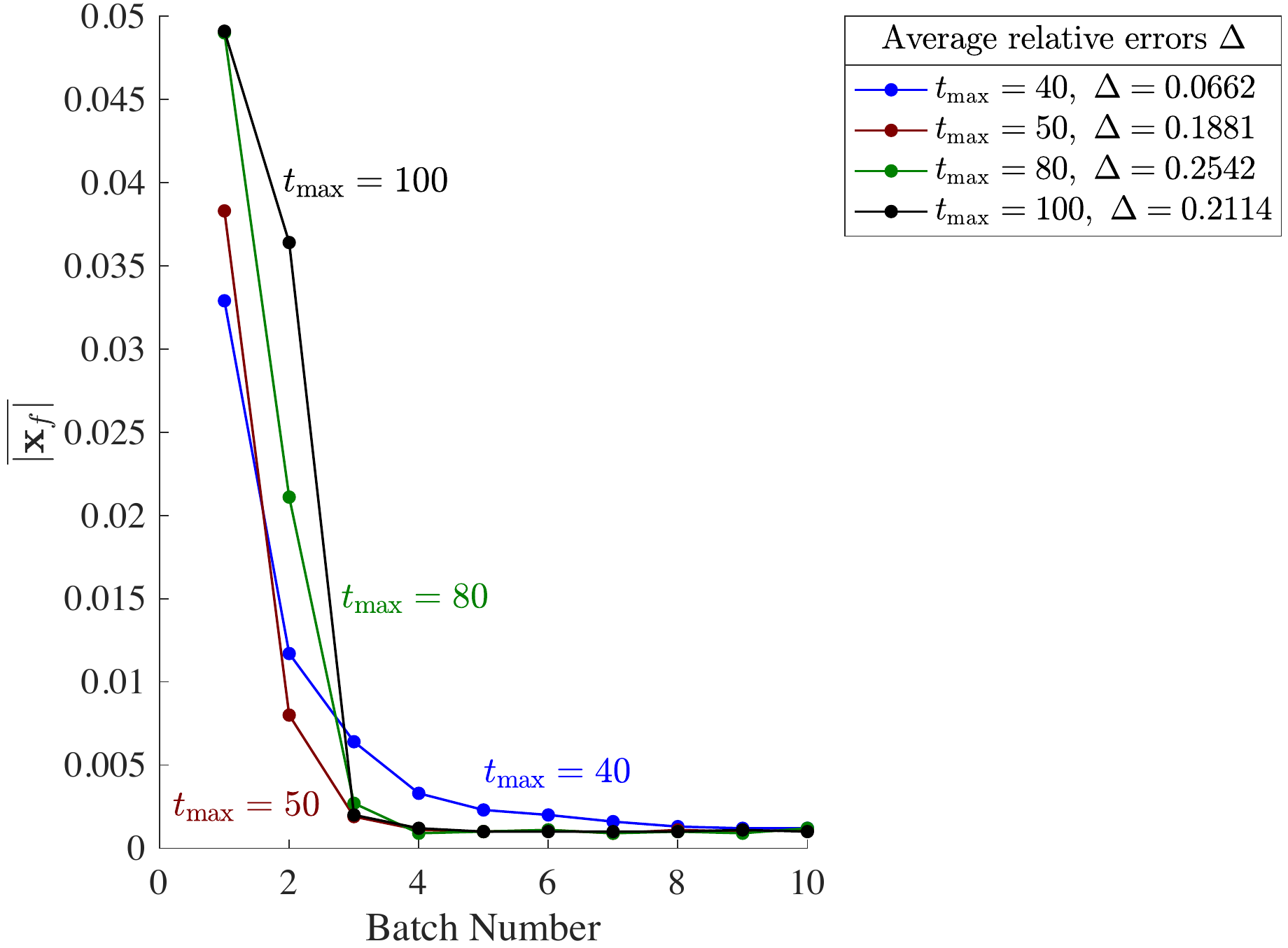}}
\caption{Average final distance of the drop from the stagnation point, $\overline{|\boldsymbol {x}_{\text{f}}|}$,  as a function of the  batch number for different values of the length of episodes $t_{\text{max}}$ (see text for the values of the other parameters). }
\label{fig:TmaxProgress}
\end{figure}

The length of the episodes $t_{\text{max}}$ is also an  important parameter. To investigate how it affects performance, we fix the values $N=4$, $\alpha=\beta=10$, $\gamma=0.95$, $p=1$ and monitor how the learning speed depends on $t_{\text{max}}$ when it is equal to {{$40$, $50$, $80$, $100$}}. For each value, we run $10$ batches of episodes of $t_{\text{max}}$ steps each, until we reach a total of {{$40,000$}} steps. This way, all batches consist of $4000$ time steps and we can compare learning speed batch by batch. The sizes of our batches are thus, respectively,  {{$100$, $80$, $50$ and $40$}}. The resulting learning curves are shown in Fig.~\ref{fig:TmaxProgress}. {{The learning curve seems to get steeper as $t_{\text{max}}$ increases, signifying that the algorithm takes longer to identify the optimal strategy}}. A possible explanation for this result is as follows. Since $\gamma$ is very close to $1$, there is very little discounting in the first few time steps. Therefore, if $t_{\text{max}}$ is large, the algorithm can afford to pick sub-optimal actions in the beginning because it has time to recover. Conversely, if $t_{\text{max}}$ is small the algorithm cannot waste time on bad actions and needs to aim for the target from the start. {{After the initial phase, the algorithm proved more accurate for larger values of $t_{\text{max}}$, likely because the drop is allowed to explore the environment for longer}}.
All learning curves consistently plateau around the $\overline{|\boldsymbol {x}_{\text{f}}|}=0.001$ mark {and relative errors are  small}. For the purpose of the experiments in Ref.~\citep{bentley1986computer}, performance is essentially the same in all four case, so we keep $t_{\text{max}}=40$, {which had the smallest $\Delta$}.

\subsection{Varying the rotation wiggle rooms}
\label{sec:varyw}

As a reminder, the wiggle room $w$ is the half-width of the window of (dimensionless) angular velocity within which  the cylinders are allowed to change their speeds. This is another important parameter  that  affects learning speed.  If the initial position $\boldsymbol x_0$ is   far from the origin,  large changes  in the fluid velocity, and therefore in the torques, may be needed to prevent the drop from wandering out of the state space. 

\begin{figure}[t]
\centering{\includegraphics[width=0.45\textwidth]{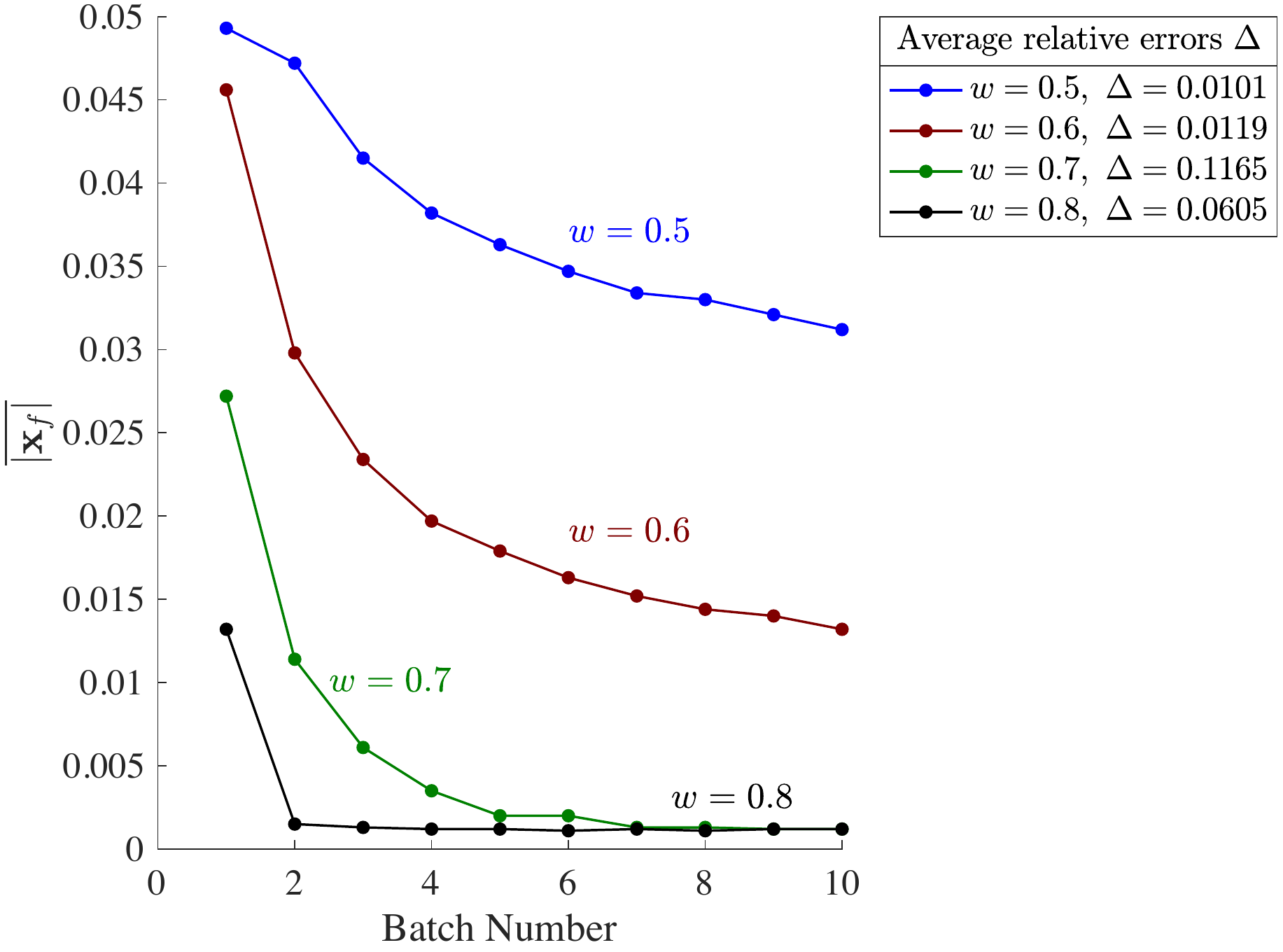}}
\caption{Average final distance of the drop from the origin, $\overline{|\boldsymbol {x}_{\text{f}}|}$,  as a function of the the batch number for four different rotation wiggle rooms {{$w$}} (see text for the values of the other parameters).}
\label{fig:LearningFailure}
\end{figure}

To illustrate this, Fig.~\ref{fig:LearningFailure} {{shows}} the learning curves obtained from $\boldsymbol x_0$ (with $\alpha=\beta=10$, $p=1$, $\gamma=0.95$, $t_{\text{max}}=40$, $N=4$) when the wiggle rooms are {{$w=0.5$, $0.6$, $0.7$ and $0.8$}}. We see that {errors are almost always negligible and that} even  a small difference in the the allowed rotations  significantly {{affects}} the learning speed; small wiggle rooms {{mean}} that we need to be more precise with our choice of actions, because we may not be able to recover from a bad one.

From a practical standpoint, a  small wiggle room might be preferable to prevent high torques and accelerations of the cylinders. However, a  value that is too small  prevents the algorithm {{from stabilising}} the drop.  In general, points further away from the origin will require bigger leeways,  and reducing the wiggle rooms decreases learning performance. {{We keep our wiggle rooms at $w=0.7$, which gave good performance while in general requiring less torque than $w=0.8$.}}

\subsection{Variance}

 \begin{figure}[t]
\centering{\includegraphics[width=0.45\textwidth]{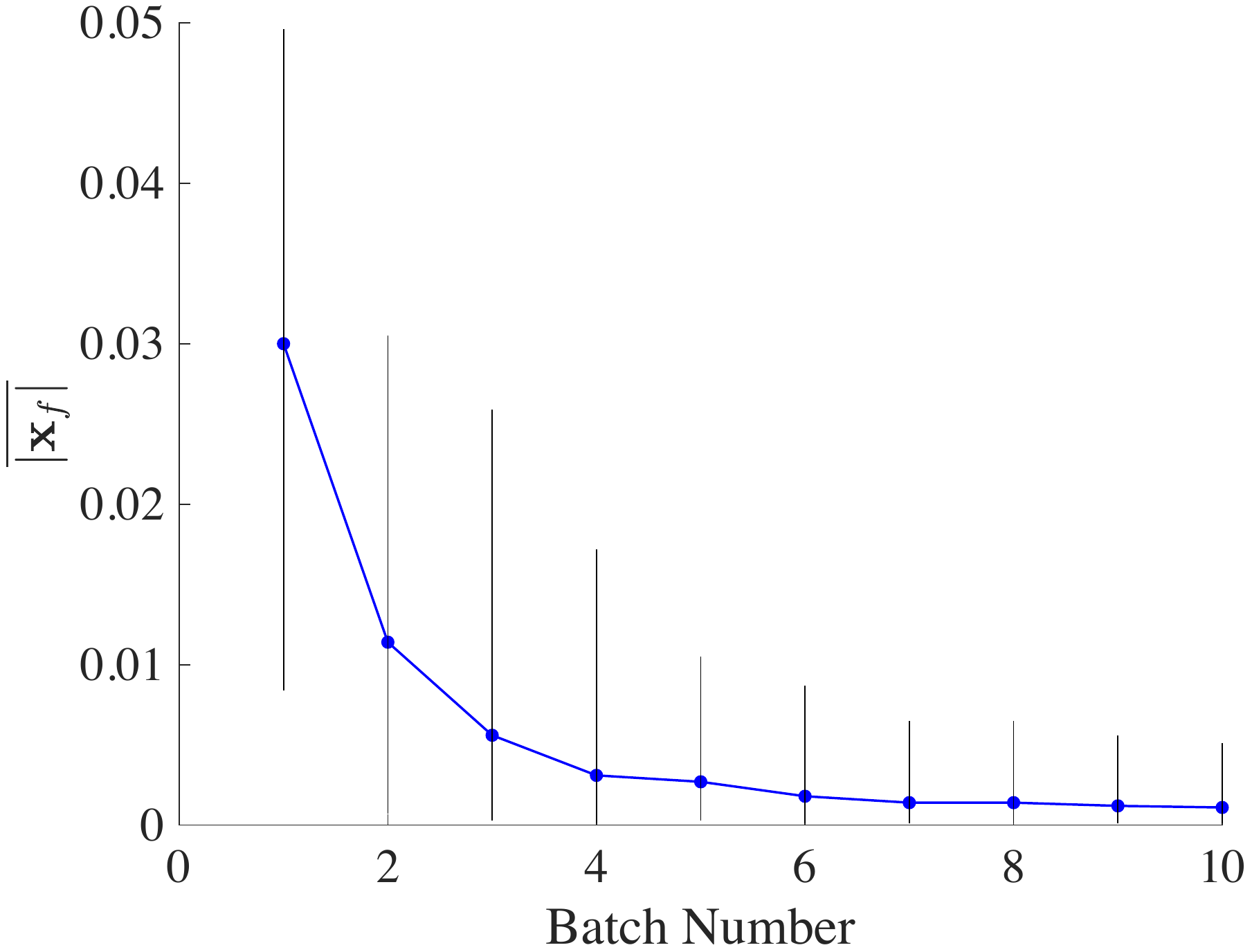}}
\caption{Variance and improvement during learning: Average final distance of the drop from the origin $\overline{|\boldsymbol {x}_{\text{f}}|}$ and range of distances as a function of the the batch number (see text for the values of the learning parameters).}
\label{fig:Spread}
\end{figure}

After running a batch, we used the resulting policy to simulate $100$ episodes in order to estimate the average final distance $\overline{|\boldsymbol {x}_{\text{f}}|}$. Let $X_i$ be the final distance from the origin in the $i$-th episode. In order for the algorithm to be useful in practice we need consistency, i.e.~for {{$\max (X_i)$ to be as small as possible. 
To test this, we run $10$ batches with the parameters $N=4$, $\alpha=\beta=10$, $\gamma=0.95$, $p=1$ and $t_{\text{max}}=40$. For each batch, in Fig.~\ref{fig:Spread} we plot $\overline{|\boldsymbol {x}_{\text{f}}|}$ along with the range of the corresponding $X_i$. We can see that the algorithm is initially rather inaccurate, but then slowly improves and becomes more consistent. In batch $10$, all episodes land within $0.0051$ of the origin, which is smaller than the (dimensionless) distance required for the experiments of Ref.~\citep{bentley1986computer}}}.

\begin{figure}[t]
\centering{\includegraphics[width=0.45\textwidth]{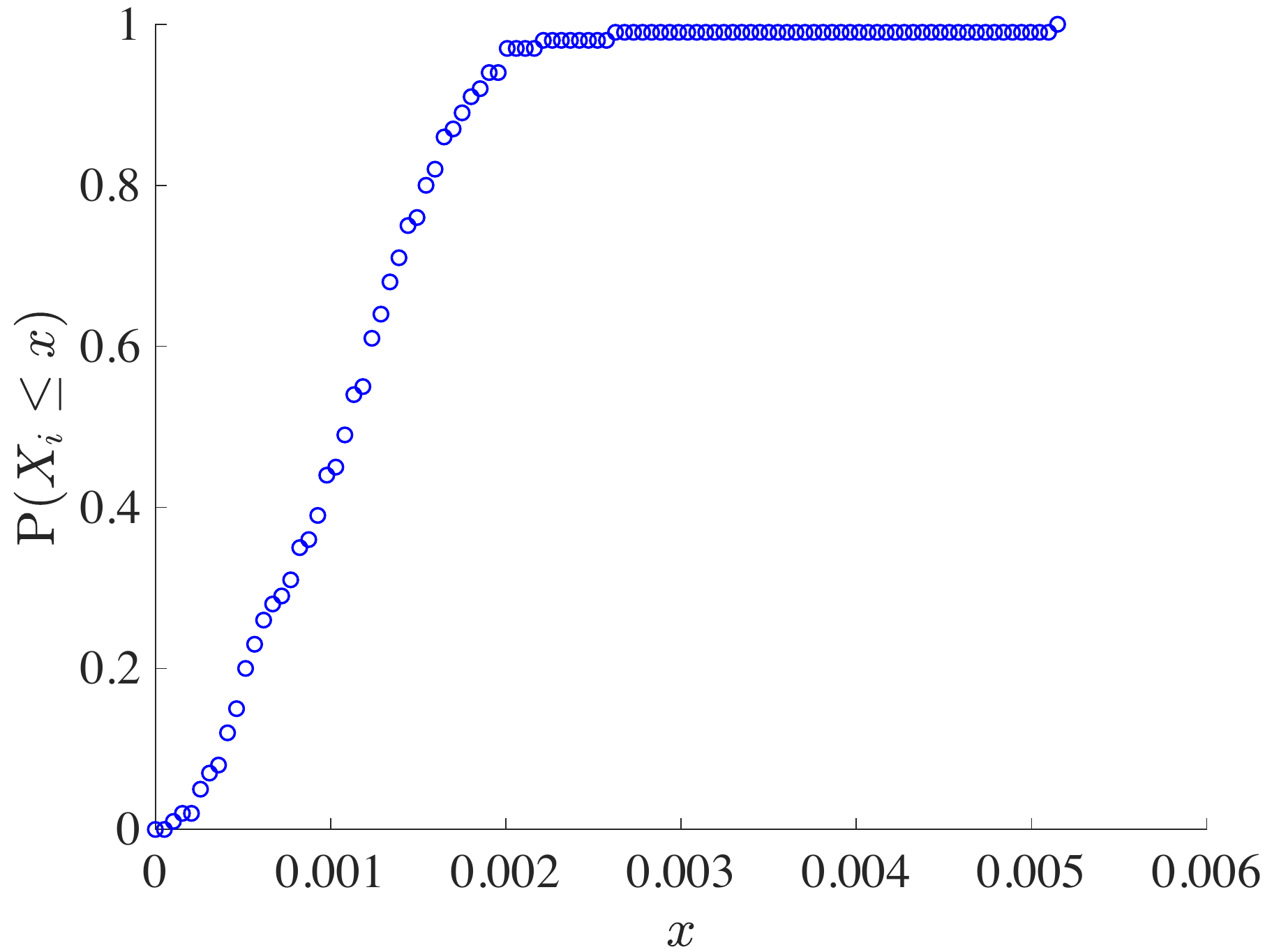}}
\caption{{{Cumulative distribution function (CDF) of the distance of the drop from the origin after $1,000$ episodes (i.e.~in batch 10; see text for the values of the learning parameters).}}}
\label{fig:CDF}
\end{figure}

 In order to better understand the distribution of $X_i$, Fig.~\ref{fig:CDF} {{shows the approximate CDF   obtained in the last batch. The distribution is clearly skewed towards $x=0$, with only $3\%$ of the $X_i$ lying in $x>0.002$.}}

\section{Robustness of the learned policy and further control}\label{Robustness}

So far we have  established the effectiveness of our Reinforcement  Learning algorithm in stabilising the drop trajectory when   the algorithm is trained against deterministic motion and when {{the drop always starts at}} a fixed location in space. In this section we relax these two  assumptions. First we establish  that the policy learned  in the absence of noise continues to work even in the presence of thermal noise (\S\ref{sec:noise}). We next study the extent to which the policy learned from a given starting point is effective when the drop starts from another location (\S\ref{sec:global}). Finally, motivated by experiments where the drop is stretched by the flow in a controlled way, we propose  a {{variant of the algorithm designed to control the}} extension rate of the flow at the location of the flow (\S\ref{sec:ratecontrol}).

\subsection{Noise}\label{sec:noise}

The dynamics of the drop  so far followed  Eq.~\eqref{dropode} with the model flow from Eq.~\eqref{eq:totalflow} and it was therefore fully deterministic. Motivated by experimental situations where the drop is small enough to be impacted by thermal noise, we now examine the performance of the deterministic Reinforcement  Learning algorithm in a noisy situation.

We    incorporate thermal noise   using a Langevin approach~\cite{batchelor1976}. In a dimensional setting, this is classically done by adding a random  term $M \boldsymbol F$ to Eq.~\eqref{dropode}  where $ M=(6\pi\mu r)^{-1}$  is the mobility of the spherical drop in {{a}} fluid of viscosity $\mu$ and  $\boldsymbol F$ is a random force. We assume that $\boldsymbol F$ has zero mean value (i.e.~$\langle   F_i(t) \rangle=0$, where we use  $\langle \rangle$ to denote ensemble averaging) {{and that it satisfies}} the fluctuation-dissipation theorem
\begin{equation}\label{eq:FD}
\langle   F_i(t)  F_j(t')\rangle= \frac{2k_BT}{M}\delta_{ij}\delta(t-t') ,
\end{equation}
where $k_B=1.3806\times10^{-23}~J\text{K}^{-1}$ is Boltzmann's constant  and $T$ is the absolute temperature.

Moving to dimensionless variables, we  used the half distance between the cylinders, $L$, as the characteristic length scale and  the inverse cylinder rotation speed, $\tau$, as the characteristic time scale (see \S\ref{sec:physicalp}), so Eq.~\eqref{eq:FD} allows to define a typical magnitude for the random force, given by $F_0= (k_B T/M\tau )^{1/2}$. Non-dimensionalising $F$ by $F_0$, the Langevin approach consists then in adding a random term of the form  $\Pe^{-1/2}\tilde F$ {{to the dimensionless version of Eq. \eqref{dropode}}}, where  $\tilde F$ is a dimensionless random  force with $\langle \tilde F_i\rangle=0$ and $\langle \tilde F_i(t)\tilde F_j(t')\rangle=2\delta_{ij}\delta(t-t')$. Here
 $\Pe$ is the dimensionless   P\'eclet number, which compares the relative magnitude {{of advection by the flow and Brownian diffusion}}
\begin{equation}\label{eq:Pe}
\Pe \equiv  \frac{L^2 }{\tau k_B T M}.
\end{equation}
We implement {{the Langevin approach}} numerically by adding a random term $\left( {2\Delta t}/{\text{Pe}}\right)^{1/2}\Gamma_i$  at the end of each numerical step, {{where $\Delta t$ is the step size used in the RK4 scheme and $\Gamma_i$ ($i=1,2$) is drawn from a standard normal distribution.}}

Physically, the P\'eclet number in Eq.~\eqref{eq:Pe} can be recast as a ratio between the radius of the drop $r$ and a thermal length scale $\ell$,
\begin{equation}\label{eq:Pe2}
\Pe =  \frac{ r  }{\ell }\quad \ell \equiv \frac{\tau k_B T}{6\pi \mu L^2}.
\end{equation}
{With the dimensions from \S\ref{sec:physicalp}, we have $L={6.35\times 10^{-2}}$~m, $\tau=8$~s, and assuming {{the fluid to be water at room temperature ($T=293$~K, viscosity $\mu= 10^{-3}$~Pas)}}, we obtain $\ell \approx 4.3\times 10^{-16}$~m. In the original  work from Ref.~\cite{bentley1986experimental}, the typical drop has radius $r\approx0.5$~mm, which leads to $\Pe\approx {1.2\times 10^{12}}$ in these experiments. This very large number clearly indicates that thermal noise was not important in this original work.}
 \begin{figure}[t]
\includegraphics[width=0.45\textwidth]{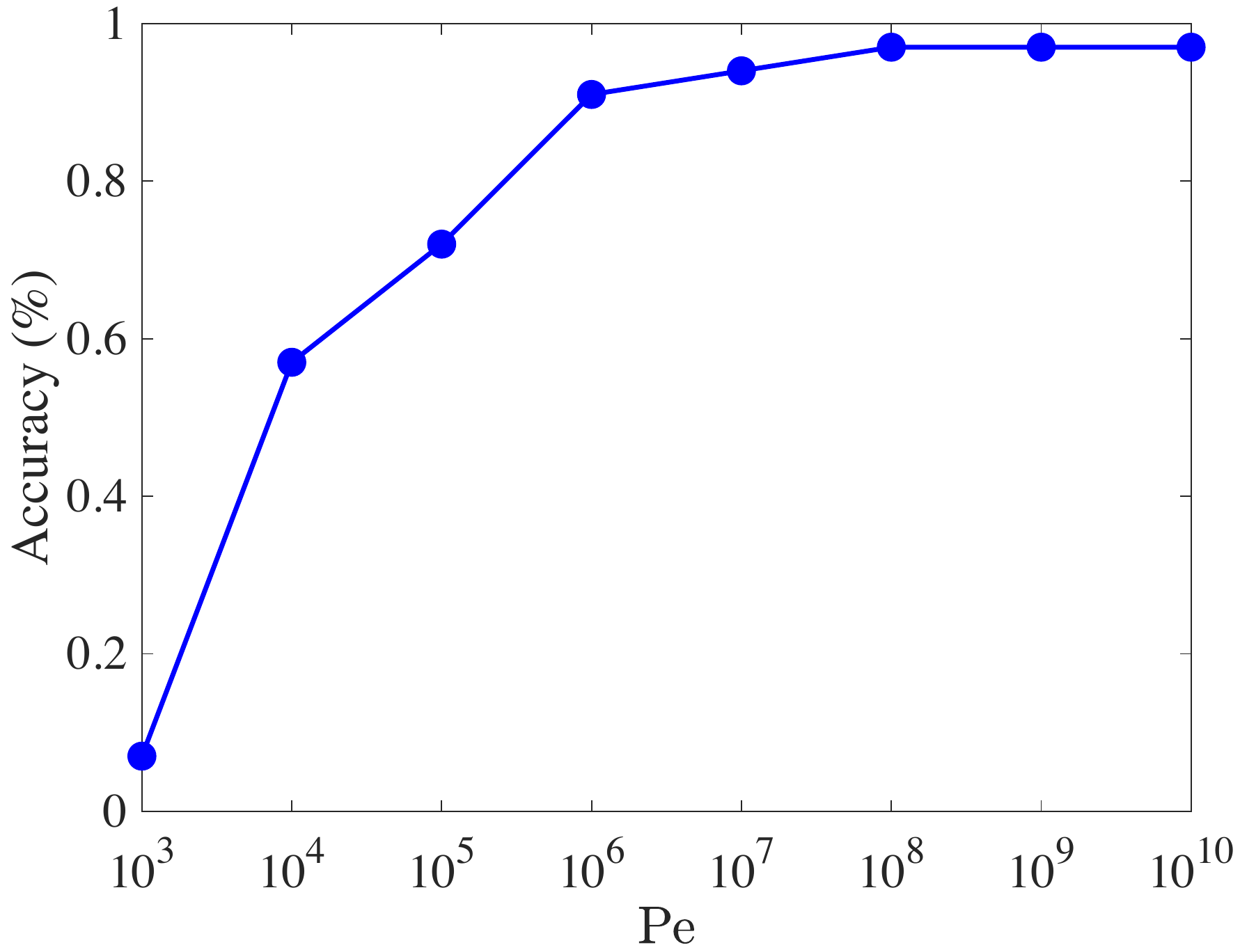}
\caption{Proportion of episodes landing within a dimensionless distance $0.0078$ of the origin for different values of the P\'eclet number ($\Pe$) using the algorithm trained in the absence of noise (see text for the values of the learning parameters). The original experiments in Ref.~\cite{bentley1986experimental}  had $\Pe\approx {1.2\times 10^{12}}$.}
\label{fig:AccuracyProgression}
\end{figure}

{{To test robustness, we applied the policy obtained via the Reinforcement Learning algorithm from the previous sections (i.e.~under deterministic drop dynamics) to environments}} with progressively smaller values of the P\'eclet number, which corresponds physically to  shrinking the  scale of the drop so that thermal noise becomes progressively more  important.   The parameters of the algorithm are {{once again}} $\alpha=\beta=10$, $N=4$, $\gamma=0.95$, $p=1$, $dt=0.025$, $t_{\rm lag}=0.0125$, $t_1=t_2=0.005$ and  $t_{\text{max}}=40$; the wiggle {{rooms are}} set to $w=0.7$.  After $500$ training runs, we let Pe take values $\mathrm{Pe}=10^k$, $3\leq k\leq 8$, and simulated $100$ {{separate}} episodes in each case. Fig.~\ref{fig:AccuracyProgression} {{shows}}  the corresponding proportions of runs landing within a distance $0.0078$ of the origin. As expected, the accuracy decreases when the Pe number becomes smaller, dropping from a largest value of $0.97$ when Pe$\Pe=10^{10}$ to $0.07$ when $\Pe=10^3$.
 The algorithm was more than $90\%$ accurate for Pe $\geq 10^6$, which is six orders of magnitude smaller than in Bentley and Leal's experiment (and thus would correspond to nanometer-sized drops). It is worth mentioning that, even though we did not do it in this work, noise could be included in the training phase rather than added once the policy has been found.

\subsection{Global Policy}\label{sec:global}

So far the learned policy was {{always obtained for the same fixed}} starting point $\x_0$. Can {{we, on the other hand, obtain}} a policy that is optimal (or sufficiently close to optimal) for {\it all} starting point? Intuitively, points in the state space that are   close together should have   similar optimal policies, so if the state space itself is sufficiently small such a global policy should exist. Should this not be the case, we would have to split the state space in smaller regions and to determine a globally good policy in each region separately.

  \begin{figure}[t]
\includegraphics[width=0.45\textwidth]{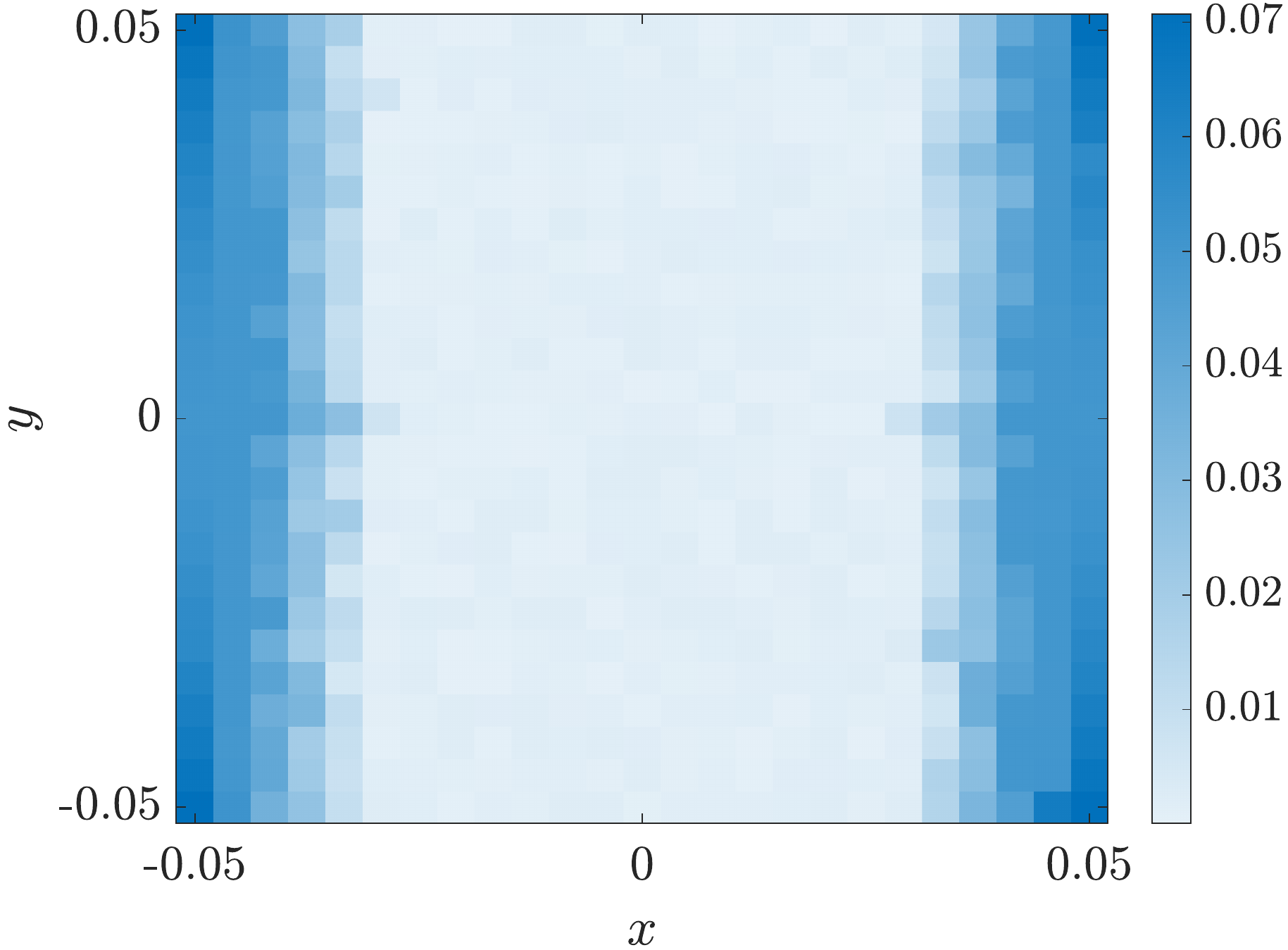}
\caption{Final distances from the origin for $625$ trajectories starting in the  $[-0.05,0.05]\times [-0.05,0.05]$ space using the   optimal policy  $\pi_*$ obtained from the single dimensionless starting position $\x_0=[-0.03,0.02]$. The time and learning parameters $dt=0.025$, $t_{\rm lag}=0.0125$, $t_1=t_2=0.005$, $\gamma=0.95$, $p=1$, $\alpha=\beta=10$, $N=4$, $t_{\text{max}}=40$. } 
\label{fig:NoiselessHeatMap}
\end{figure}

\begin{figure}[t]
\includegraphics[width=0.45\textwidth]{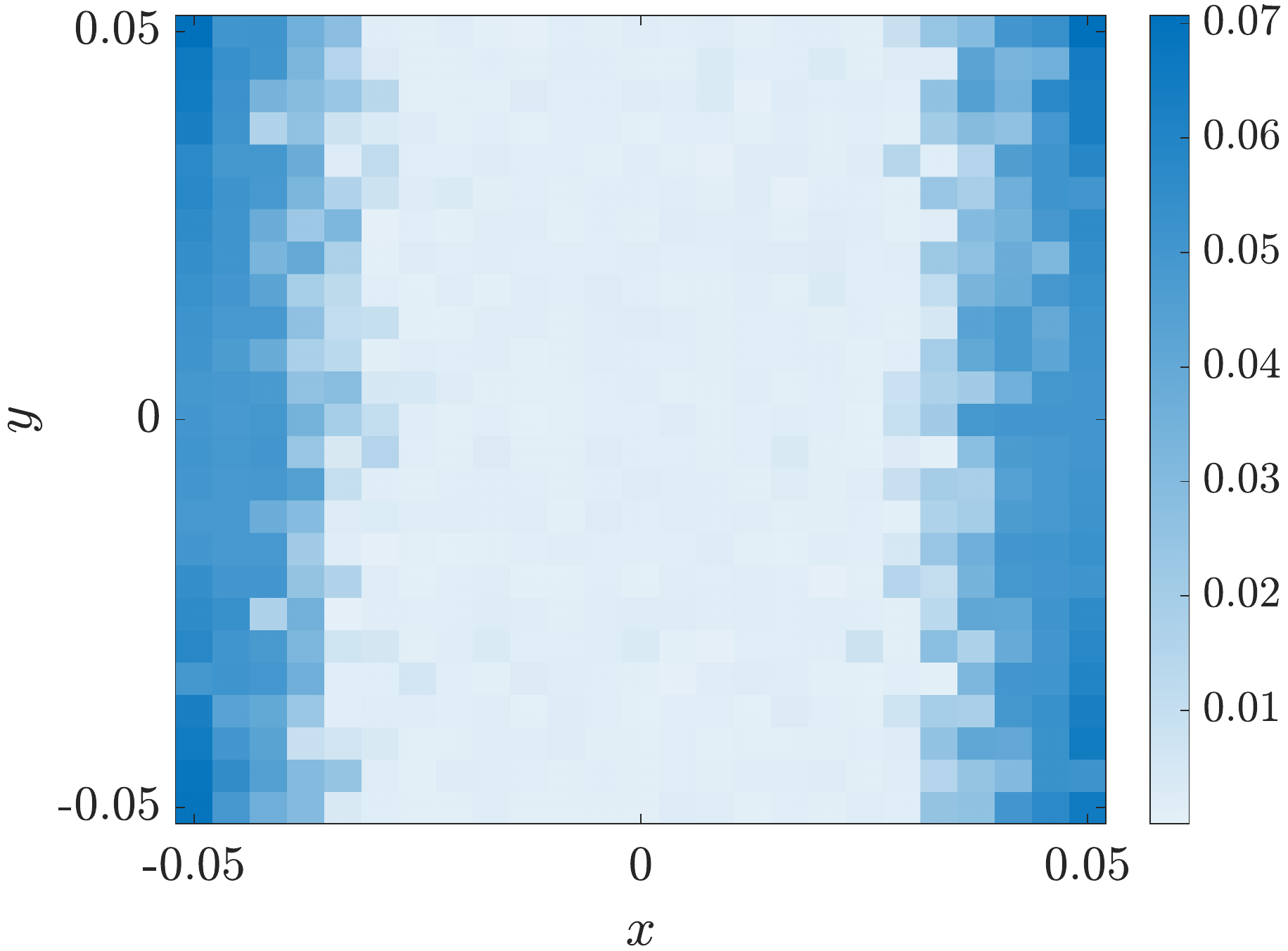}
\caption{Same as Fig.~\ref{fig:NoiselessHeatMap} in the case where thermal noise is added to the drop trajectory with  {{$\Pe=10^5$}}.} 
\label{fig:HeatMapNoise}
\end{figure}

We investigate the existence of a global policy by   estimating    the optimal policy $\pi_*$ from the dimensionless starting position $\x_0=[-0.03,0.02]$ and then running  trajectories from a number of other points in the state space.  We use the time and learning parameters  $dt=0.025$, $t_{\rm lag}=0.0125$, {{$t_1=t_2=0.005$}}, $\gamma=0.95$, $p=1$, $\alpha=\beta=10$, $N=4$, $t_{\text{max}}=40$. After $500$ training runs in a noiseless environment, we construct a $25\times 25$ rectangular lattice of evenly spaced points in the state space  $[-0.05,0.05]\times [-0.05,0.05]$ and run a trajectory from each one of them using the policy   obtained for $\boldsymbol x_0$ (i.e.~no further learning occurs during that process). We then use the results to build a colour map of the final distances, i.e.~a $25\times 25$ matrix $M$ where $M_{ij}$ is coloured according to the final distance from the origin  of a trajectory starting from  {{the corresponding}} location on the grid. We show the results for all $625$ trajectories in the absence of  thermal noise in  Fig.~\ref{fig:NoiselessHeatMap}. We can see  that all unsuccessful starting points are clustered around the edge of the state space on the sides where the flow points away from the origin, suggesting that we can indeed find a global policy by making the state space a bit smaller. The algorithm was successful in {{$61.12\%$}} of cases, with an average final distance of $0.0160$ and a standard deviation of $0.0213$. The average final distance is heavily skewed by the edge cases. In the region $[-0.03,0.03]\times [-0.05,0.05]$, corresponding the lighter strip in the middle, the average final distance was {{$0.0012$}} with a {{standard deviation}} of {{$7.7026\times 10^{-4}$ and a success rate of $100\%$}}. The largest final distance {{in this region}} was {{$0.0075$}} and the smallest was {{$6.1237\times10^{-5}$}}. To see how thermal noise affects this result, we carry out the same simulations by incorporating noise as in \S\ref{sec:noise} in the case where  {{$\Pe=10^5$}}, with results shown in Fig.~\ref{fig:HeatMapNoise}. The algorithm was now successful in {{$60.96\%$}} of cases, with an average final distance of {{$0.0164$}} ad a standard deviation of {{$0.0209$}}. Again, if we restrict the set of initial states to those in $[-0.03,0.03]\times [-0.05,0.05]$ these figures improve significantly. Success rate jumps to {{$97.07\%$}} and the average final distance becomes {{$0.0019$}} with a standard deviation of {{$0.0020$}}. The largest final distance {{in this region}} was {{$0.0153$}} and the smallest was {{$1.4041\times10^{-4}$}}. In summary, the overall performance was quite similar to the noiseless case, except for a small decrease in accuracy and consistency in the central region. This shows that the algorithm is robust to noise even in the case of nanometer-sized drops.

\subsection{Extension Rate Control}\label{sec:ratecontrol}

Returning to the  physical aspects of the experiment,  Taylor's  original study addressed how the properties of the flow affected the shape of the drop~\cite{taylor1934formation}.  When the drop is fixed at the origin, its rate of deformation is dictated by the eigenvalues of the velocity gradient tensor, $\nabla\boldsymbol u(\boldsymbol 0)$. 
The flows considered in this study are two-dimensional and irrotational {{so}} that $\nabla\boldsymbol u$ remains symmetric and traceless throughout. {{The velocity gradient}} is {{thus}} characterised by a pair of eigenvalues $\lambda_0>0$ (\textit{extension} rate) and $-\lambda_0< $  (\textit{compression}). When we alter the  speeds of the cylinder with the control algorithm, we inevitably change the eigenvalues of $\nabla\boldsymbol u(\boldsymbol x)$, where $\boldsymbol x$ is the position of the  drop, leading to a time-dependent eigenvalue $\lambda(t)$. Since this eigenvalue controls the deformation of the drop, we wish to keep its magnitude as close as possible to the extension rate $\lambda_0$ which we aim to study while we control   the drop position.

\begin{figure*}[t]
\includegraphics[width=0.95\textwidth]{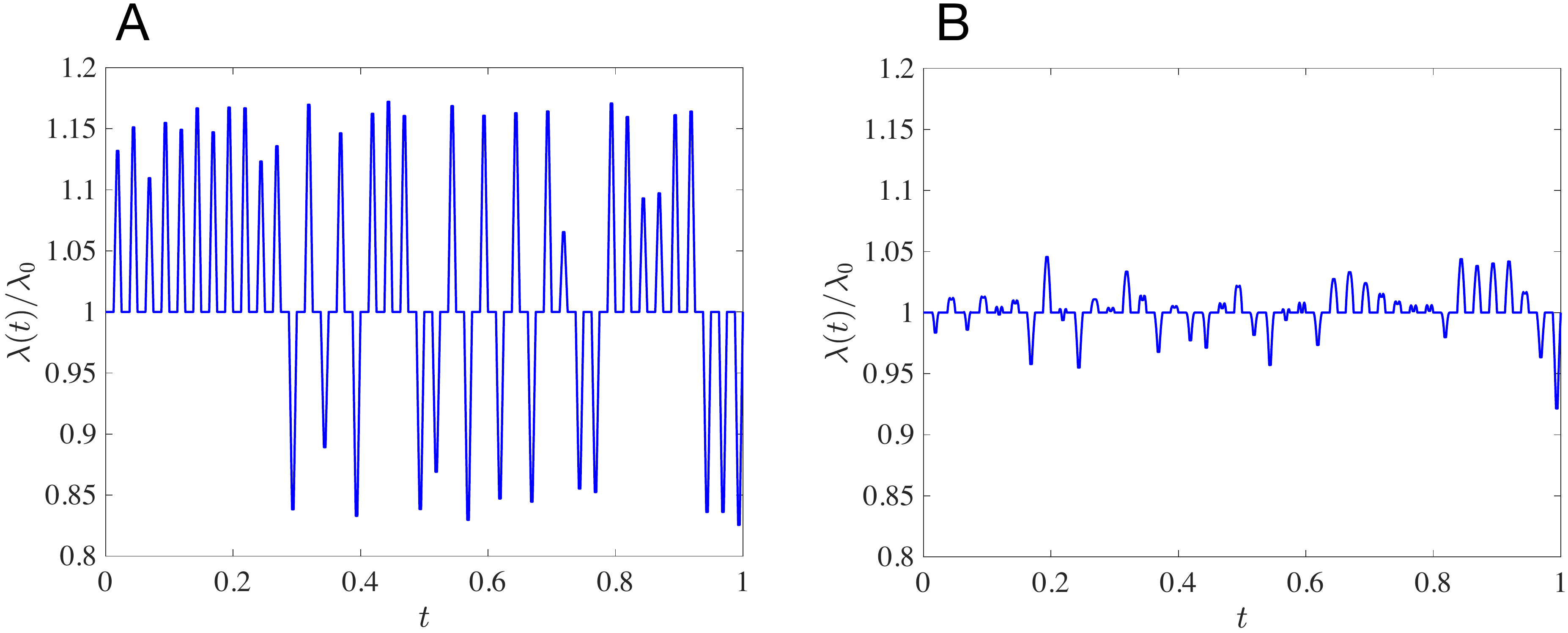}
\caption{Ratio between the extension rate $\lambda(t)$ and the reference value $\lambda_0$ at the centre of the uncontrolled apparatus.  
A: Without scaling, the extension rate routinely varies by $ 15\%$ around $\lambda_0$. 
B:  Using the piecewise linear scaling from Eq.~\eqref{eq:scaling}, the fluctuations of the  extension rate near the drop are significantly reduced. 
The time and learning parameters are $dt=0.025$, $t_{\rm lag}=0.0125$, $dt=t_1=t_2=0.005$, $\gamma=0.95$, $p=1$, $\alpha=\beta=10$, $N=4$, $t_{\text{max}}=40$ while the P\'eclet number is 
{{$\Pe=10^5$}}.
} 
\label{fig:ExtensionRates}
\end{figure*}

Here we examine the case where the starting position is $\boldsymbol x_0=[-0.03,0.02]$, with the same parameters as above (i.e.~$dt=0.025$, $t_{\rm lag}=0.0125$, $dt=t_1=t_2=0.005$, $\gamma=0.95$, $p=1$, $\alpha=\beta=10$, $N=4$, $t_{\text{max}}=40$). We assume the drop is subject to thermal noise with $\Pe=10^5$. After $500$ training runs, we simulate a $40$-step trajectory (in which no learning occurs) during which we sample the extension rate $20$ times per time step. 
In Fig.~\ref{fig:ExtensionRates}A we plot the variation  of the scaled extension rate, {{$\lambda(t)/\lambda_0$, with time, where $\lambda_0$ is the value at the centre of the uncontrolled apparatus}} (using the dimensionless parameters in the problem, we have $\lambda_0=1.28$). The extension rate is seen to undergo significant variations during the controlled motion of the drop, with jumps that are routinely $\pm 15\%$ about the desired value $\lambda_0$. The norm of the final state was $0.0056$.

To lower the variations on $\lambda(t)$ and keep it closer to its target value, we modified the algorithm as follows. The idea is to note that if an angular velocity vector $\boldsymbol \Omega$ (i.e.~the vector of all four cylinder rotations) induces an extension rate $\lambda$ at $\boldsymbol x$, then by linearity the angular velocity {{vector}} $k\boldsymbol \Omega$   induces an extension rate $k\lambda$ at the same point ($k\in\mathbb R$).  
We may then   scale, at each time step,  $\boldsymbol \Omega$ with a suitable scalar function $s(t)$ so that   the angular velocity vector changes as $s(t)\cdot\boldsymbol \Omega(t)$, where $\boldsymbol \Omega(t)$ corresponds to standard speed modulation. Since the jumps in $\lambda(t)$ are due to the rapid changes in angular velocities, we choose $s(t)$ to minimize the impact of speed modulation. Specifically, at time step $t_k$ and state $\boldsymbol x_k$ we denote $\Lambda=\lambda_0/\hat\lambda$, where $\hat\lambda$ is the extension rate at $\boldsymbol x_k$ {{resulting from unscaled speed modulation}}. Then we take a piece-wise linear scaling
\begin{equation}\label{eq:scaling}
s(t)=
\begin{cases}
1,\, t_k\leq t\leq t_k+t_{\rm lag},\\
\left|(\Lambda-1)\frac{t-(t_k+t_{\rm lag})}{\delta}+1\right|,\, 0\leq t-(t_k+t_{\rm lag})\leq \delta,\\
\Lambda \ \ \ \text{ if } t_{\rm lag}+\delta\leq t-t_k \leq dt-\delta,\\
\left|(1-\Lambda)\frac{t-(t_k+dt)+\delta}{\delta}+\Lambda\right|,\,-\delta\leq t-(t_k+dt)\leq 0,
\end{cases}
\end{equation}
{{and choose $\delta=t_1=t_2$}}. To compensate for this scaling, we also make the change $dt\to dt/\Lambda$. {{In}} Fig.~\ref{fig:ExtensionRates}B {{we show}} the evolution of the extension rate (scaled by $\lambda_0$) in a trajectory with the same parameters as before but with our scaling implemented. {{A couple of large excursions remain}}, but performance has noticeably improved relative to the original control algorithm  (left). {{The norm of the final state was $9.4389\times10^{-4}$, indicating that scaling does not affect accuracy}}. This proof-of-principle result shows therefore that a scaling in the optimal policy can be used to limit the extension rate in the flow.

 \section{Discussion}

In this paper we saw how Reinforcement Learning can be applied to solve a classical control problem for fluid dynamics at low Reynolds numbers. Our goal here was to modulate the rotation speeds of a model of Taylor's four-roll mill  in  order to stabilize a drop positioned near the stagnation point, which is known   to be unstable. We implemented an Actor-Critic method and found a probabilistic policy that worked well for all initial positions.

In our approach, we proceeded by steps. We first derived  a basic version of the algorithm, and then  added  measurement delays, thermal noise and extension rate control. The algorithm was able to manoeuvre the drop effectively in all cases and the  accuracy achieved was below that required in the experiments of Bentley and Leal~\cite{bentley1986experimental}, and therefore  satisfactory for most  experimental implementations.  {Numerical results shown in \S \ref{sec:results} also demonstrated that learning is remarkably consistent, with minimal variance within the learning curves in the majority of cases.}

The good performance observed was, to a large extent, due to our choice of actions rather than to the quality of the approximation for the policy  ($\pi$). Indeed, numerical results in  Fig.~\ref{fig:PerformanceVsN} show that a first order approximation of the form $\pi\approx K\exp(ax+by)$ is sufficient to get accurate results. In practice, the learning process was often slower at the beginning, when the algorithm had not yet gathered enough information to take good actions. Then, once the general shape of the policy had been identified, learning sped up significantly, until it slowly tapered off as we approached the theoretical accuracy.

 It is worth mentioning here  that we  attempted other implementations  too, which were not successful . We initially tried to discretize the state space, so that the drop would move in a finite grid as opposed to a continuous environment, but it was difficult to combine this with the Markov property and  harder to factor in thermal noise. We also  used a truncated Fourier series for the form of the function $f$, but this was  computationally expensive and  it artificially introduced discontinuities as well as Gibbs' phenomena. 
 
 Finally, we also experimented with the shape of the reward function, seeking to penalise actions requiring very large torques. {Unfortunately,  our attempted modifications in that regard (such as subtracting some simple increasing function of the torque from $R_t$) did not succeed.} After extensive simulations, we concluded that torque {reduction} can be achieved by either shrinking the sate space (so that smaller corrections are needed) or reducing the default angular velocities.

There are many possible extensions to our work. One could try an algorithm with higher sample efficiency (i.e.~one which makes better use of past experience), one that better balances exploration and exploitation, or a different learning paradigm altogether (i.e.~a neural network). We may also implement different approximations for the value function as well as alternative rewards and sampling methods. It would also be interesting to devise a model where time is continuous.

From a physical standpoint, it might be desirable to include inertia  (both  of the drop and the fluid) and include a nonzero response time to variations in $\boldsymbol \Omega$. We could also  allow ourselves to act on more than one cylinder at a time or to undo an action by exploiting the time-reversibility of the viscous flow. Another area for improvement is extension rate control, since some jumps in Fig.~\ref{fig:ExtensionRates}  still remain. 
Finally, one could devise a model where one only has incomplete knowledge of the drop's position.

\section*{Acknowledgements}
This project has received funding from the European Research Council (ERC) under the European Union's Horizon 2020 research and innovation programme  (grant agreement 682754).

\section*{Supplementary Material}

We include in Supplementary Material~\cite{SM} a movie of the trajectory displayed in 
Fig.~\ref{fig:illustrate}. In the movie, the histogram on the left shows the current angular velocity of each cylinder, as well as their average value. The diagram on the right shows the motion of the drop inside the state space as well as the {rotation} of each cylinder (note that the  radii of both the drop and the cylinders are not to scale) {and the eigenvectors of $\nabla\u$ at the location of the drop (note that since the flow is irrotational, these are also the eigenvectors of the rate-of-strain tensor)}. For clarity, the cylinders are displayed on the corners of the state space, rather than in their actual locations.

\section*{Matlab code}

The code created in this work is freely available as a matlab .m file on GitHub~\cite{GH}. To estimate the optimal policy, the user needs to initialize the parameters, add a section break as indicated and run as many batches (outer "for" loops) as needed. The  parameter \texttt{AverageDistance}  corresponds to the average final distance for the current batch, and can be used to assess performance. By commenting out lines $76-82$ in the code, the program can be used to simulate trajectories during which no learning occurs.

\appendix

\section{Proof of rejection sampling algorithm} \label{sec:proof}
We need to show that the conditional distribution of $X$ is $p$. Let $P$ be the cumulative distribution function of $p$ and $Q$ that  of $q$. Then by Bayes theorem
\begin{equation}
\mathbb P\left(X\leq x|Y\leq \alpha\right)=\frac{\mathbb P\left(Y\leq \alpha|X\leq x\right)Q(x)}{\mathbb{P}(Y\leq \alpha)}
\end{equation}

\begin{align}
\mathbb P\left(Y\leq \alpha|X\leq x\right) &=\frac{\mathbb P\left(Y\leq \alpha, X\leq x \right)}{Q(x)}\\
&=\int^x \frac{\mathbb P(Y\leq \alpha|X=t)}{Q(x)}q(t)\mathrm dt\\
&=\frac{1}{Q(x)}\int^x\frac{A\cdot p(t)}{B\cdot q(t)}q(t)\mathrm dt\\
&=\frac{A\cdot P(x)}{B\cdot Q(x)}
\end{align}
Also 
\begin{align}
\mathbb{P}(Y\leq \alpha)=\int_I\frac{A\cdot p(t)}{B\cdot q(t)}q(t)\mathrm dt=\frac{A}{B}
\end{align}
Substituting, we see that $\mathbb P\left(X\leq x|Y\leq \alpha\right)=P(x)$, so, conditional on being accepted, $X\sim p$.\\

\bibliographystyle{unsrt}

\bibliography{Marco_references}

\end{document}